\documentclass[prb,reprint,amssymb,amsmath]{revtex4-2}
\usepackage{graphicx}
\usepackage{subcaption}
\usepackage{color}
\newcommand\thsnd[1]{#1\thinspace000}

\begin{document}
\title{Graph-based machine learning beyond stable materials and relaxed crystal structures}
\author{Filip Ekström}
\affiliation{Division of Statistics and Machine Learning, Department of Computer and Information Science (IDA), Linköping University}
\author{Rickard Armiento}
\affiliation{Theoretical Physics, Department of Physics, Chemistry and Biology (IFM), Linköping University}
\author{Fredrik Lindsten}
\affiliation{Division of Statistics and Machine Learning, Department of Computer and Information Science (IDA), Linköping University}

\date{\today}

\begin{abstract}
There has been a recent surge of interest in using machine learning to approximate density functional theory (DFT) in materials science. However, many of the most performant models are evaluated on large databases of computed properties of, primarily, materials with precise atomic coordinates available, and which have been experimentally synthesized, i.e., which are thermodynamically stable or metastable. These aspects provide challenges when applying such models on theoretical candidate materials, for example for materials discovery, where the coordinates are not known. To extend the scope of this methodology, we investigate the performance of the Crystal Graph Convolutional Neural Network (CGCNN) on a data set of theoretical structures in three related ternary phase diagrams (Ti,Zr,Hf)-Zn-N, which thus include many highly unstable structures. We then investigate the impact on the performance of using atomic positions that are only partially relaxed into local energy minima. We also explore options for improving the performance in these scenarios by transfer learning, either from models trained on a large database of mostly stable systems, or a different but related phase diagram. Models pre-trained on stable materials do not significantly improve performance, but models trained on similar data transfer very well. We demonstrate how our findings can be utilized to generate phase diagrams with a major reduction in computational effort.
\end{abstract}

\maketitle

\section{Introduction} 
Discovering new materials is a driving force for new technologies. With the increase in computational resources, there has been a surge in available data from automated high-throughput materials simulations, obtained using supercomputers. The computed properties have been made available in online databases where one can search for materials with specific desired properties, which enables new applications \cite{armiento_database-driven_2020}. 

A central concern in the design of a new material is its thermodynamical stability. This is determined by the formation energy of a phase in relation to that of other phases at the same composition, as well as the most stable combination of competing phases it can decompose into. Hence, an extensive, in principle exhaustive, search of the corresponding chemical space is needed to properly determine stability. A common computational method for this purpose is density functional theory (DFT) \cite{hohenberg_inhomogeneous_1964-1, kohn_self-consistent_1965-1}, which on the semi-local level of theory is usually considered as having a fairly low computational cost compared to other quantum-mechanics-based methods, but nevertheless limits how many materials that can be screened. Using machine learning as an approximation of conventional simulation methods such as DFT has recently gained interest \cite{schmidt_recent_2019, schleder_dft_2019, chen_critical_2020}. Many methods are reported with impressive results with prediction errors close to or lower than that of DFT \cite{faber_prediction_2017}, and their low computational effort make them highly interesting for complementing or replacing quantum-mechanical-based calculations for, e.g., computing formation energies. 

Much effort has been put into finding suitable representations, so called \emph{descriptors}, of crystal structures for use in machine learning methods. Early works include, for example, smooth overlap of atomic positions \cite{bartok_representing_2013}, three descriptors inspired from the Coloumb matrix description of molecules \cite{faber_crystal_2015}, descriptors based only on stoichiometry \cite{faber_machine_2016}
, and the many-body tensor-representation \cite{huo_unified_2018}.

An alternative approach to predicting material properties from a descriptor of the crystal structure is to learn an interatomic potential (see, e.g., Refs.\ \onlinecite{szlachta_accuracy_2014, dragoni_achieving_2018, shapeev_moment_2016, podryabinkin_active_2017}). The use of interatomic potentials are perhaps more associated with molecular dynamics simulations, but they can also be used to relax the atomic positions and predict the formation energies of competing phases in a phase diagram \cite{gubaev_accelerating_2019}.

Recently, much work has been directed towards so called \emph{feature learning} which aims to find machine learning models that can learn \emph{both} suitable descriptors \emph{and} the target value from these descriptors. In particular, the use of neural networks for graph data, so called graph neural networks (GNNs), have received more interest in chemistry and physics. One motivation for this is that molecules can naturally be described as graphs \cite{duvenaud_convolutional_2015, kearnes_molecular_2016}, but a pioneering work by \textcite{xie_crystal_2018} shows that this type of model is also very appealing for crystals. \citeauthor{xie_crystal_2018} demonstrate the predictive performance of their Crystal Graph Convolutional Neural Network (CGCNN) by predicting multiple properties of crystals, including the formation energies of 9350 crystals from the Materials Project database \cite{Jain2013}. A number of works have followed (see, e.g., Refs.\ \onlinecite{sanyal_mt-cgcnn_2018,chen_graph_2019, park_developing_2020}) which all develop GNNs for crystal structures and which are evaluated using data from the Materials Project. 

However, most of the available databases of computational materials properties, such as the Materials Project, are based on structures found experimentally, e.g., as reported in the Inorganic Crystal Structure Database \cite{bergerhoff_inorganic_1983-1} or Crystallography Open Database \cite{Downs2003, grazulis_crystallography_2009, grazulis_crystallography_2012}. This gives a bias towards stable, or close to stable, structures, meaning evaluation of models on these databases is confined to a restricted subspace of all possible phases. Hence, the performance for generated, possibly highly unstable, theoretical structures cannot be assumed to match this performance.

Candidate materials for a phase diagram are typically created by substituting atoms into the crystal structures of other, already known, materials. The atomic positions of these candidate materials can be far away from the local energy minima. If one seeks the formation energy of the relaxed structure, the model has to, in principle, perform two tasks. Not only does it have to predict the formation energy, but it needs to do so by overcoming the unrelaxed structure. It has been shown that this issue indeed affects the predictive performance of GNNs, which are relying on knowledge about the structure \cite{park_developing_2020, noh_uncertainty-quantified_2020}. 

The performance of machine learning methods on highly unstable material phases, as needed to predict phase diagrams, has so far not been considered in great detail. Neither has it been thoroughly studied how the predictive performance of a model is affected when predicting formation energies of structures which are not completely relaxed, and how it is improved when the structure moves closer to the relaxed geometry. To investigate this, we perform an empirical evaluation, using the CGCNN model and a data set with theoretical structures from three related ternary phase diagrams, Hf-Zn-N, Ti-Zn-N, and Zr-Zn-N generated by \textcite{tholander_strong_2016}. First, we investigate the performance when CGCNN is trained with this data set of mostly unstable materials. Then, we investigate how the performance changes when the atomic positions have only been partially relaxed into local minima. When the performance in these settings has been established, we explore options for reducing the need for training data by using transfer learning. This is done by pre-training a model using either a large database with a variety of mostly stable materials, or data from one of the related phase diagrams. The pre-trained models are then fine-tuned with data from the phase diagram of interest.

We conclude our findings by using CGCNN to speed up the generation of phase diagrams. CGCNN predicts the formation energies of partially relaxed structures between steps in the relaxation process, and structures far from the convex hull are discarded and excluded from further relaxation. Thus, the total computational cost is reduced.

\section{Background}
\subsection{Composition phase diagrams and thermodynamical stability \label{sec:pdsconvexhulls}}
A composition phase diagram shows the stable phase of a material as a function of its composition. A phase is stable if its formation energy is lower than any linear combination of competing phases. This means that the stability of each phase can be determined by constructing a convex hull of the phases in the phase diagram \cite{ong_lifepo2_2008, ong_thermal_2010}. If a phase belongs to the convex hull it is stable, otherwise, it will decompose into the linear combination of phases on the hull that decreases the total energy. To generate a phase diagram without any prior knowledge of which phases that are stable, a large number of potential materials needs to be considered since one in principle need to determine the formation energy of every possible phase at every composition.

This type of phase diagrams are highly relevant in materials design since finding materials that are stable is a central aspect. Additionally, they solve the crystal structure prediction problem, i.e., for a given composition, the corresponding crystal structure can be determined.

Computational methods, in particular DFT, are often used to predict formation energies of unknown phases. To generate the potential materials, other elements are substituted into already known materials \cite{kirklin_open_2015,hautier_data_2011}. This procedure generates hypothetical materials for which the exact geometry is unknown. Hence, before their formation energy can be calculated, they have to go through a relaxation procedure which typically consist of a number of steps where the cell volume, cell shape and atom positions are changed into local minimum of the formation energy. 

\subsection{Graph neural networks \label{sec:gnns}}
\begin{figure}
\begin{subfigure}{0.50\linewidth}
\includegraphics[width=\textwidth]{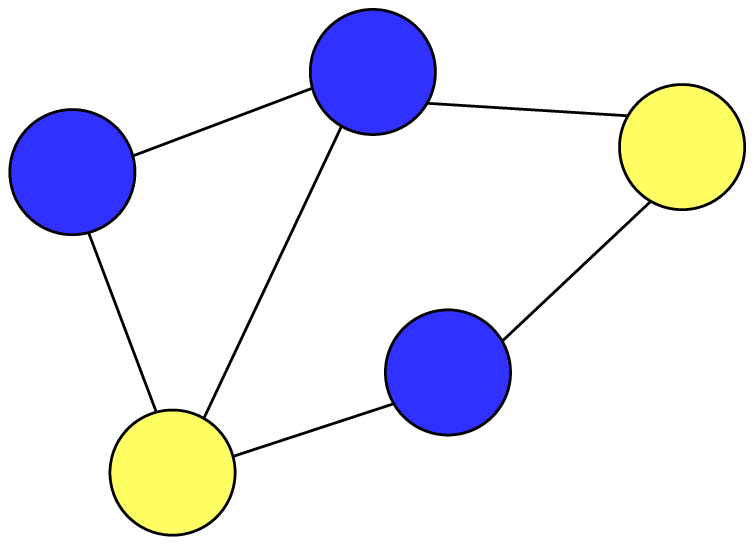}
\caption{\label{fig:gnn_illustration_before}}
\end{subfigure}%
\hfill
\begin{subfigure}{0.5\linewidth}
\includegraphics[width=\textwidth]{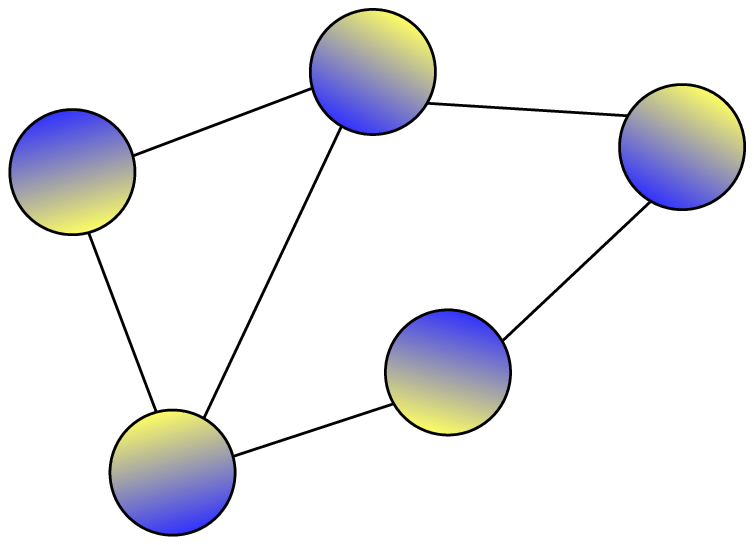}
\caption{\label{fig:gnn_illustration_after}}
\end{subfigure}%
\caption{\label{fig:gnn_illustration} Illustration of how the hidden node representations are updated. Information is aggregated from the neighborhood of each node, meaning that after some iterations, the node feature vector contain information about its initial state, but also about its neighborhood. (a) Initial graph (b) after $t$ iterations. Normally, a node is represented by a high dimensional vector, and the colors are just for illustrative purposes.}
\end{figure}

Graph neural networks (GNNs) are a special kind of artificial neural networks designed to handle graph structured data. Let $G$ be a graph defined by $(V,E)$ where $V$ is the set of nodes and $E$ is the set of edges. A node $v \in V$ is associated with a feature vector $\mathbf{x}_v$ and an edge $(v,w) \in E$ with a feature vector $\mathbf{e}_{vw}$. A GNN is then a parametrized function $f_{\theta}(\cdot)$ that computes some target value $y=f_{\theta}(\mathbf{x})$, where $\mathbf{x}  = \{ \mathbf{x}_v: v \in V\} \cup \{ \mathbf{e}_{vw} : (v,w) \in E \}$. In short, the function $f_{\theta}(\cdot)$ is based on a message passing scheme where a hidden representation $\mathbf{h}_v$ of a node $v \in V$ is iteratively updated by aggregating information from its neighborhood, with each iteration corresponding to a layer in the neural network. Following the notation introduced by \textcite{gilmer_neural_2017}, the hidden representation of a node $v \in V$ after iteration $t$ is called $\mathbf{h}_v^t$. It is updated by first computing a message $\mathbf{m}_v^{t+1}$ as
\begin{subequations}
\begin{equation}
\mathbf{m}_v^{t+1} = \sum_{w \in N(v)} M_t(\mathbf{h}_v^t,\mathbf{h}_w^t,\mathbf{e}_{vw}) \label{eq:gnn_m}
\end{equation}
and then computing node representation as
\begin{equation} 
\mathbf{h}_v^{t+1} = U_t(\mathbf{h}_v^t,\mathbf{m}_v^{t+1}).
\end{equation}
\end{subequations}
Here, $N(v)$ represents the neighborhood of $v$, and the message function $M_t(\cdot)$ and update function $U_t(\cdot)$ are parametrized by $\theta$ and learned via, e.g., stochastic gradient descent.
The process of propagating information to the neighbourhood is illustrated in Fig.~\ref{fig:gnn_illustration}.

In this work we use the Crystal Graph Convolutional Neural Network (CGCNN) \cite{xie_crystal_2018}. A crystal is modeled from the unit cell as a graph with the atoms as nodes and edges between the closest neighboring atoms within a certain distance, taking into account the periodic boundary. The initial atom representation $\mathbf{x}_v$ is a 92-dimensional vector with one-hot encoded physical properties of the atom, and the vectors $\mathbf{e}_{vw}$ encoding the distance between atoms, expanded with a radial basis function similar to SchNet~\cite{schutt_schnet_2018}.

\subsection{Ensembles and uncertainty quantification}
An ensemble combines multiple models by averaging the predictions made by the individual ensemble members into a final prediction. One reason to use ensembles is that it has been shown to improve the predictive performance \cite{lakshminarayanan_simple_2017}. Furthermore, it provides a way of estimating the total uncertainty of the predictions. This can be divided into \emph{aleatoric} and \emph{epistemic} uncertainty. Aleatoric refers to the underlying natural randomness. In machine learning terms, it is the uncertainty that remains even if the model parameters $\theta$ are perfectly known. Epistemic instead refers to uncertainty due to lack of knowledge. This arises in machine learning since a model is trained from a finite training data set, resulting
in (epistemic) uncertainty about the parameter values themselves. For example, an ensemble could consist of $N$ different neural networks, trained on the same data but with random initializations and stochastic optimization. Thus, $N$ plausible parameter configurations $\{\theta_i\}_{i=1}^N$ are obtained, which gives $N$ different predictions. Using the spread of the predictions of the ensemble members has proven to be an efficient way of quantifying epistemic uncertainty for neural networks \cite{NEURIPS2019_8558cb40, lakshminarayanan_simple_2017}.

\subsection{Data sets}
In this work we use a data set of \textcite{tholander_strong_2016} (TAATA, after the name of the authors). This data set contains DFT calculations, including structure relaxation, for a large number of materials phases in the phase diagrams Hf-Zn-N, Ti-Zn-N, and Zr-Zn-N. These three phase diagrams make up the three subsets TAATA-Hf, TAATA-Ti, and TAATA-Zr. The three subsets consist of 4839 (TAATA-Hf), 4328 (TAATA-Ti), and 3650 (TAATA-Zr) materials, each with three different relaxation steps: preprerelax, prerelax and final relax, with final relax being the most accurate level. These relaxations were performed using the projector augmented wave method (PAW) \cite{blochl_projector_1994} and Perdew-Burke-Ernhofer generalized gradient approximation (PBE-GGA) \cite{perdew_generalized_1996} as exchange-correlation functional, implemented in the Vienna \textit{Ab-initio} Simulation Package \cite{kresse_efficiency_1996, kresse_efficient_1996} (VASP) (v5.4.1). All calculations used a 600 eV plane wave energy cutoff. In the first relaxation step, preprerelax, only the unit cell volume is relaxed, which requires on average approximately 5 core minutes/phase. In the prerelax and final relaxation steps, volume, cell shape, and the internal degrees of freedom describing the atomic coordinates are relaxed, with the final step done with higher precision (the VASP-setting PREC set to "Accurate" instead of "Normal", which controls the number of grid points in the Fourier transform grids and the accuracy of the PAW projector representation in real space). These two steps require approximately 60 core minutes/phase (prerelax) and 540 core minutes/phase (final).

We have also used a model trained on a data set extracted from Materials Project by \textcite{xie_crystal_2018} (the model can be found together with the CGCNN code \footnote{The code and the pre-trained model are from \protect\url{https://github.com/txie-93/cgcnn}}). This training data set consists of roughly \thsnd{37} materials with completely relaxed structures, covering 87 elements, and each material consisting of up to seven different elements. Apart from covering a different chemical space, this data set has not been produced with the same computational parameters as TAATA.

\subsection{Transfer learning}
A machine learning model requires data from DFT calculations to learn to approximate DFT, and typically improves with more data. This means that there is a trade-off between the computational expense of producing DFT simulations (training data) and model accuracy. Relying on training data that, in some sense, is expensive to obtain is common in machine learning (e.g., images annotated by humans according to what they depict), and much effort has therefore been put into investigating methods to reduce the need for data. One such method is transfer learning \cite{yosinski_how_2014}. The idea is simple: a model is pre-trained on one problem, after which it is fine-tuned by training on data from a different, but similar, problem. This can be seen from the perspective of feature learning as assuming that the two problems could be represented well with more or less the same features. If the problem of interest is suffering from low availability of training data, these features can be learned better from a different problem where more data is available, and then transferred to the new problem domain.

\section{\label{sec:method}Method}
We use an ensemble of five ($N=5$) CGCNNs. Each network is trained on the different subsets of the TAATA data set using an L1 loss function and the Adam optimizer \cite{kingma_adam_2017}. Similarly to the original work, the optimized hyperparameters are the learning rate, number of layers, and weight decay, using a grid search with completely relaxed structures from all three TAATA subsets. The combination of hyperparameters with the best MAE on a validation set was then chosen for all subsequent investigations. For the other hyperparameters, the default values provided in the CGCNN code \cite{Note1} were used.

The individual ensemble members are trained with access to the same training data. However, some of the training data is reserved to be used only as validation data to avoid overfitting by early stopping. To capture the uncertainty of model parameters (i.e., the epistemic uncertainty), the split of training/validation data is different for the different members, in addition to initializing the model parameters randomly and using stochastic-gradient-based training. When using CGCNN to aid in the generation of a phase diagram (described later in this section), we estimate the total uncertainty as $\sigma = \sqrt{\sigma_a^2 + \sigma_e^2}$, where $\sigma_a^2$ and $\sigma_e^2$ are the aleatoric and epistemic uncertainty respectively. We estimate $\sigma_a^2$ as the average of the mean squared errors (MSE) of the ensemble members when predicting formation energies of the validation set during training, and $\sigma_e^2$ as the variance of the predictions made by the individual ensemble members.

To investigate transfer learning, we have compared two different scenarios. In the first scenario, data is assumed to be available for a phase diagram that is related to an unknown phase diagram of interest. W want to use this data to improve the predictive performance for the unknown phase diagram. This situation could appear, for example, when a study is expanded into a larger chemical space. As a second scenario, related phase diagrams are not available, but we instead try to utilize data from a large database of materials involving all elements like Materials Project. We explore and compare both these scenarios by (i) transferring a model pre-trained on one TAATA phase diagram by fine-tuning with data of another; (ii) transferring a model trained by \textcite{xie_crystal_2018} on Materials Project data (available online \cite{Note1}) by fine-tuning with data from one of the TAATA phase diagrams.

Finally, we illustrate how our findings can be used to accelerate the generation of a phase diagram with CGCNN using the following scheme: Start by creating a set of candidate structures, $S$, e.g., by substituting desired chemical elements in structure prototypes available in present materials databases. This set is divided into a training set $S_{\text{train}}$ and an evaluation set $S_{\text{eval}}$. The structures in $S_{\text{train}}$ are fully relaxed, through the prepre-, pre-, and final relaxation steps, and their formation energies are obtained. Using the partially relaxed training structures and their final formation energies, two (ensembles of) CGCNN models are trained for the preprerelax and prerelax steps, respectively. Then, all the structures in $S_{\text{eval}}$ are relaxed to just the preprerelax step. The preprerelax-level CGCNN model predicts the final formation energies of the partially relaxed evaluation structures and a phase diagram is constructed based on these predictions together with all the structures in the training set. Then, all structures in $S_{\text{eval}}$ which, in the predicted phase diagram, has an energy above the resulting convex hull of stability $E_{\text{hull}} > 2\sigma$ are removed (with $\sigma$ being the total uncertainty of the prediction described above). The remaining structures go through the prerelax step and the procedure of predicting a phase diagram and removing structures far from the convex hull is repeated. The remaining structures are then relaxed through the final step and the phase diagram is constructed using the energies computed by DFT for the structures in $S_{\text{train}}$ and the non-discarded structures in $S_{\text{eval}}$. 

\section{Results}
Test sets for evaluating the performence were created by setting aside 20~\% of the materials for each TAATA subset. For the partially relaxed structures, structures used for training and testing have been relaxed equally far. However, the target formation energy is always the final formation energy, obtained at the last relaxation step.

\subsection{Training on different relaxation levels}
\begin{figure}
\centering
\includegraphics[width=\columnwidth]{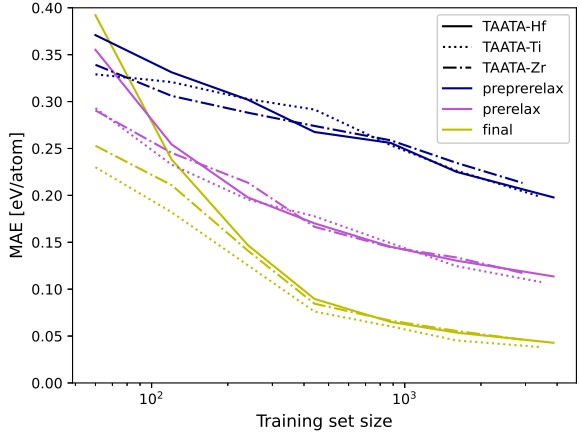}
\caption{\label{fig:relax_levels} Test MAE for different relaxation levels and phase diagrams in the TAATA data set. There are no major differences between the three phase diagrams, and the errors for the different levels of relaxation are ordered as expected.}
\end{figure}
Test MAEs for models trained from scratch are shown in Fig.~\ref{fig:relax_levels}. For the largest training set size, i.e., all of the data points not used for testing (2920, 3460, and 3870 materials for TAATA-Zr, TAATA-Ti, and TAATA-Hf, respectively) the test MAE in eV/atom is 0.038 for TAATA-Ti, 0.043 for TAATA-Hf, and 0.046 for TAATA-Zr when using fully relaxed structures. The corresponding test errors for prerelax is 0.11~eV/atom (TAATA-Hf and TAATA-Ti) and 0.12~eV/atom (TAATA-Zr), and for preprerelax 0.20~eV/atom (TAATA-Hf and TAATA-Ti) and 0.21~eV/atom (TAATA-Zr). It is clear that the performance improves with the amount of training data, and in general worsen with more inaccurate descriptions of the atomic positions, as expected. 

Comparing these results with using the sine-based Coloumb matrix together with kernel ridge regression (KRR) \cite{faber_crystal_2015}, the Coloumb matrix seems to be very robust with respect to the relaxation levels. For example, using the largest training size of 3870 training materials from the TAATA-Hf data set, the test MAE is 0.33~eV/atom for preprerelax, 0.32~eV/atom for prerelax and 0.31~eV/atom for the final relaxation level. However, these errors are still much larger than those of the CGCNN model. It also seems like the Coloumb matrix with KRR suffers from overfitting for small training sizes with a test MAE of more than 0.9~eV/atom for TAATA-Hf when using 60 training datapoints (for full details on the results with KRR, see supplemental material \footnote{See Supplemental Material at [URL will be inserted by publisher] for full details on the results}). To put the results in some perspective, a model always predicting a formation energy of 0 eV/atom would yield a test MAE of roughly 0.5~eV/atom. (A model that instead predicts according to the mean of the largest training set, roughly $-0.2$~eV/atom for all three subsets, improves that MAE with less than 0.05~eV/atom.)

\begin{figure}
\includegraphics[width=\columnwidth]{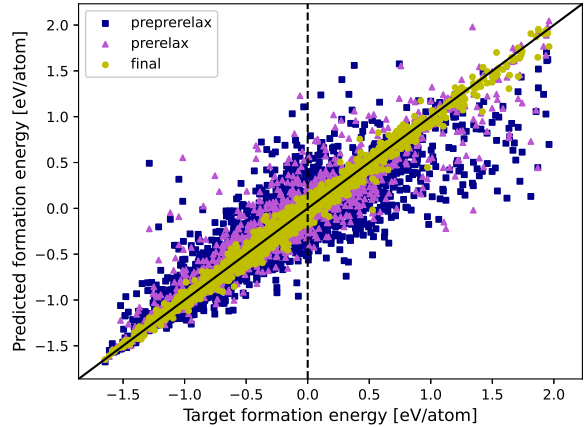}
\caption{\label{fig:pred_vs_actual_scratch} Predicted vs actual formation energy for models trained with all data except for the test data (3870, 3460, and 2920 datapoints for TAATA-Hf, TAATA-Ti, and TAATA-Zr respectively). The overall slope of the points vs. the ideal line (black) suggests a tendency to underestimate the formation energy at higher values, and underestimate at lower values.}
\end{figure}

We further investigate how the predicted formation energies compare to the actual formation energies in the predicted-versus-target scatter plot in Fig.~\ref{fig:pred_vs_actual_scratch} (since the performance is similar between the different phase diagrams, we have only separated the predictions according to relaxation level and not according to phase diagram). Apart from again demonstrating that more accurate structures give better predictions (as can be seen also in Fig.~\ref{fig:relax_levels}), Fig.~\ref{fig:pred_vs_actual_scratch} shows a small tendency of the model to underestimate high formation energies and overestimate lower formation energies for the less accurate relaxation levels. 

\begin{figure}
\includegraphics[width=\columnwidth]{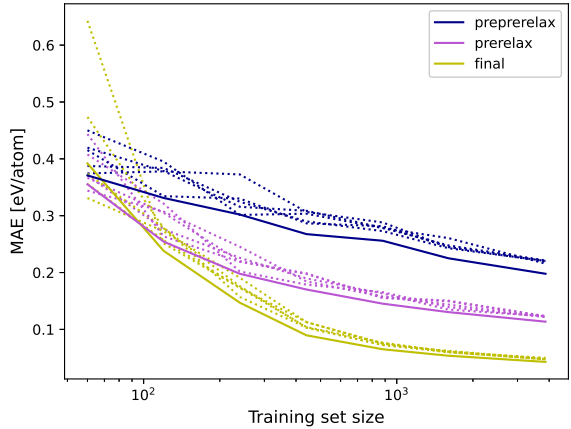}
\caption{\label{fig:ensemble_plot} Test MAE for the ensemble (solid lines) compared with the individual members (dashed lines) for the TAATA-Hf data set. The ensemble consistently gives a lower MAE than the individual ensemble members, except at the smallest training set sizes.}
\end{figure}

All predictions using CGCNN have, so far, been the average of the five ensemble members. To demonstrate the improved performance enabled by using an ensemble, we compare the test MAE of the ensemble with those of the individual ensemble members, using the TAATA-Hf data set. The results can be seen in Fig.~\ref{fig:ensemble_plot}, and demonstrate the power of using ensembles; the MAE of the ensemble is consistently lower than that of the individual members, except for the smaller training sets where individual members occasionally outperform the ensemble. On the other hand, in the case of small training sets, some of the members are also far worse. We obtain the same qualitative results for the other data sets; see the supplemental material \cite{Note2}. 

\subsection{Transfer learning}
\begin{figure*}
\centering
\includegraphics[width=\textwidth]{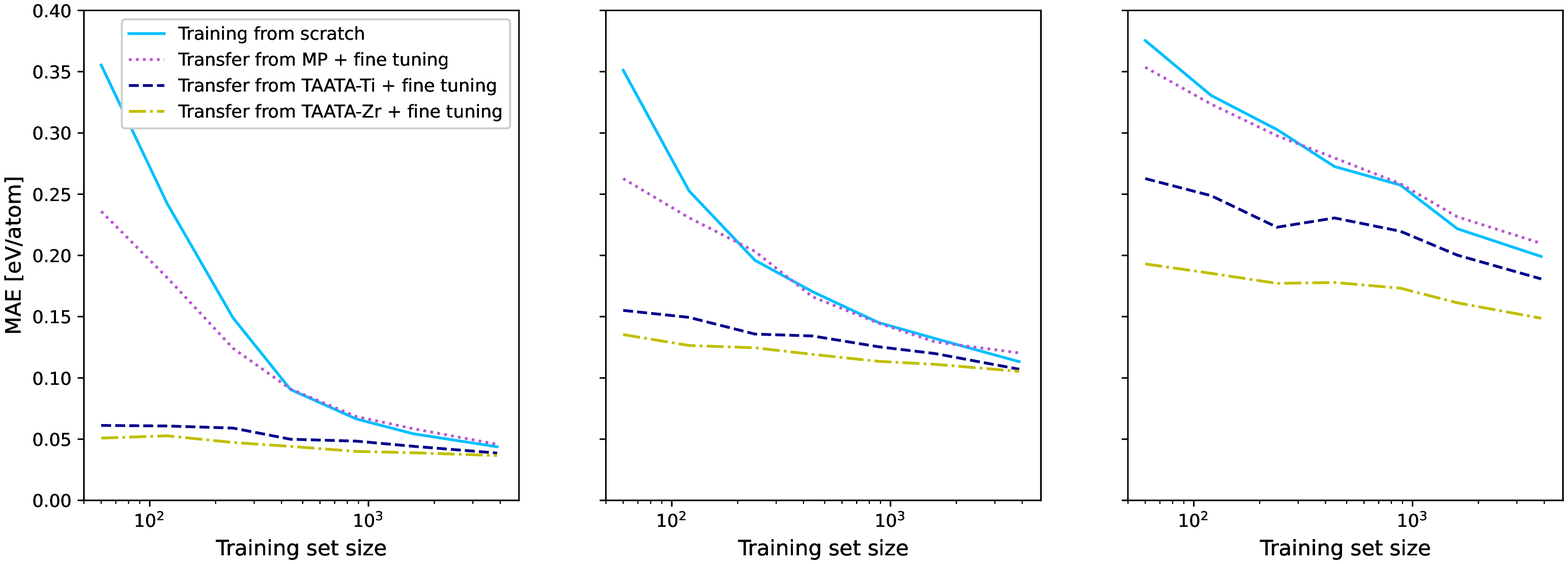}
\caption{\label{fig:transfer} Test MAE when transferring different pre-trained models to the TAATA-Hf data set for relaxation levels final (left), prerelax (middle), and preprerelax (right) and comparing it with training a model from scratch. Training set size corresponds to the number of data points used for fine-tuning and does not include training data for pre-training.}
\end{figure*}

Test errors when transferring models to the TAATA-Hf data set are shown in Fig.~\ref{fig:transfer}. Apart from pre-training on a different TAATA subset, we have also used a pre-trained CGCNN model provided by \textcite{xie_crystal_2018}, trained on data from Materials Project (MP) (available online \cite{Note1}). The results when training a model from scratch are provided as a comparison. For the pre-trained models, test MAEs before being fine-tuned are presented in Table~\ref{tab:hf_pretrain}. 
\begin{table}
\begin{tabular}{ cccc }
 & final & prerelax & preprerelax  \\
 \cline{2-4}
MP & 0.58 & 0.56 & 0.49  \\ 
TAATA-Ti & 0.77 & 0.34 & 0.29  \\ 
TAATA-Zr & 0.49 & 0.50 & 0.38  \\ 
\end{tabular}
\caption{\label{tab:hf_pretrain} Test MAEs (eV/atom) for models pre-trained on different data sets when predicting formation energies of materials from TAATA-Hf. The relaxation levels indicate which structures have been used for fine-tuning (and hence, testing). TAATA models are pre-trained on the same relaxation levels as transferred to, whereas the MP models are transferred from the same model.}
\end{table}
To put the poor performance of the pre-trained models without fine-tuning into perspective, we reiterate that simply guessing the formation energy as 0 (alternatively, as the mean of the training data) gives a test MAE of roughly 0.5 eV/atom. However, as shown in Fig.~\ref{fig:transfer}, as soon as the models are fine-tuned on a small training data set from the actual phase diagram of interest, the errors drop significantly. This is particularly apparent for the models pre-trained on the related TAATA data set. 

\begin{figure*}
\begin{subfigure}{0.33\linewidth}
\includegraphics[width=\textwidth]{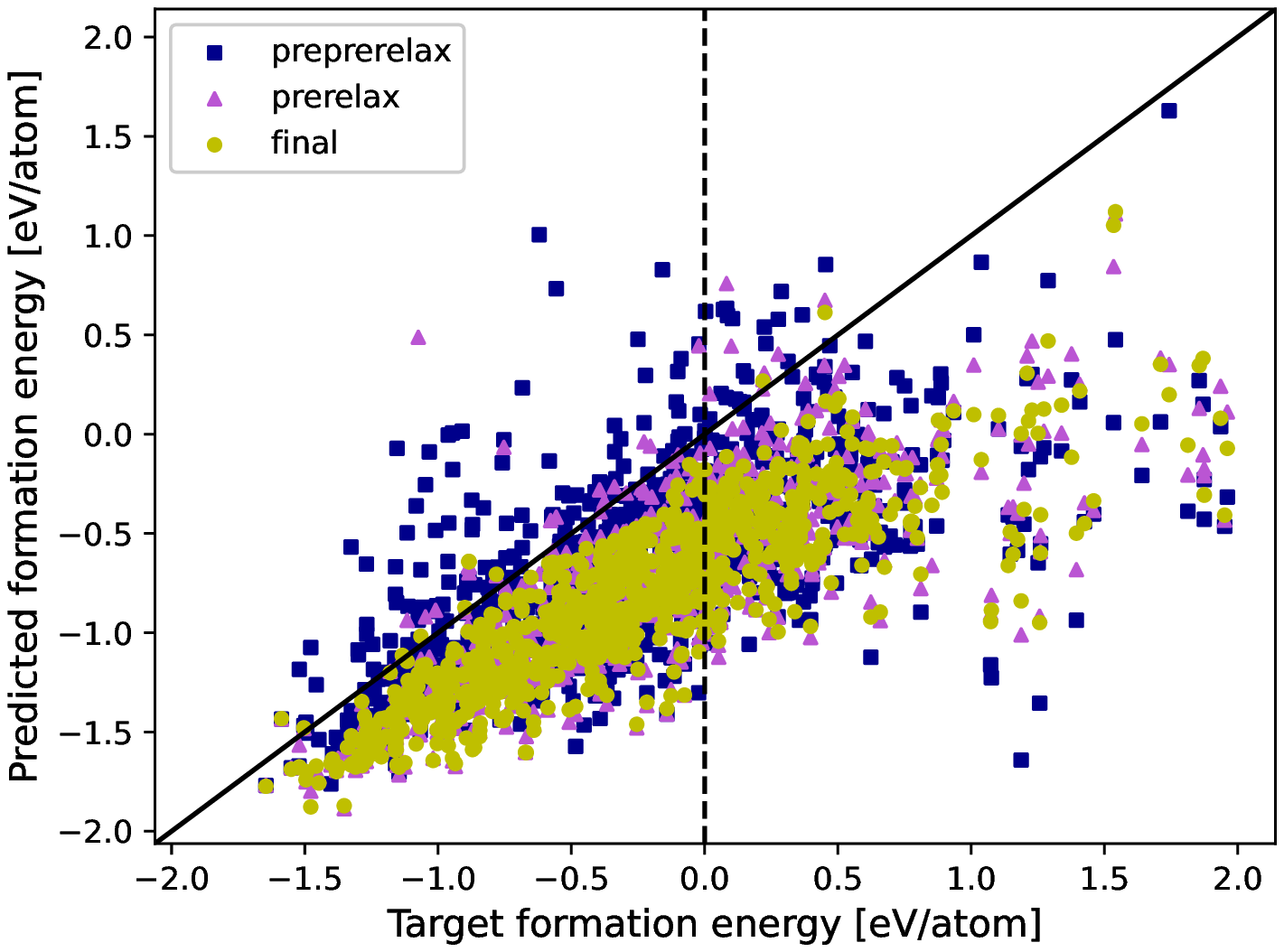}
\caption{\label{fig:xie_pretrained_pred_vs_actual_no_finetuning}}
\end{subfigure}%
\hfill
\begin{subfigure}{0.33\linewidth}
\includegraphics[width=\textwidth]{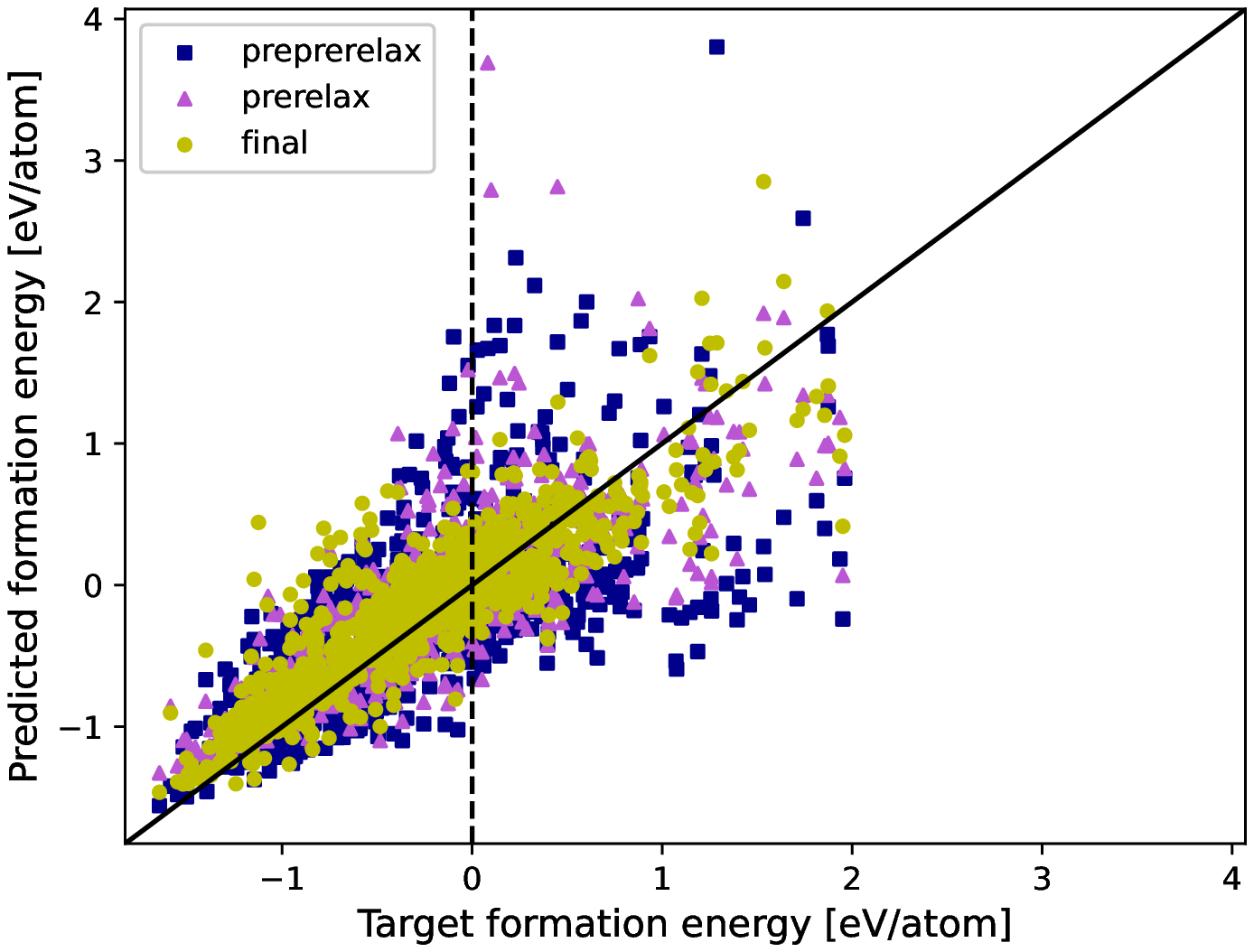}
\caption{\label{fig:xie_pretrained_pred_vs_actual_60_finetuning}}
\end{subfigure}%
\hfill
\begin{subfigure}{0.33\linewidth}
\includegraphics[width=\textwidth]{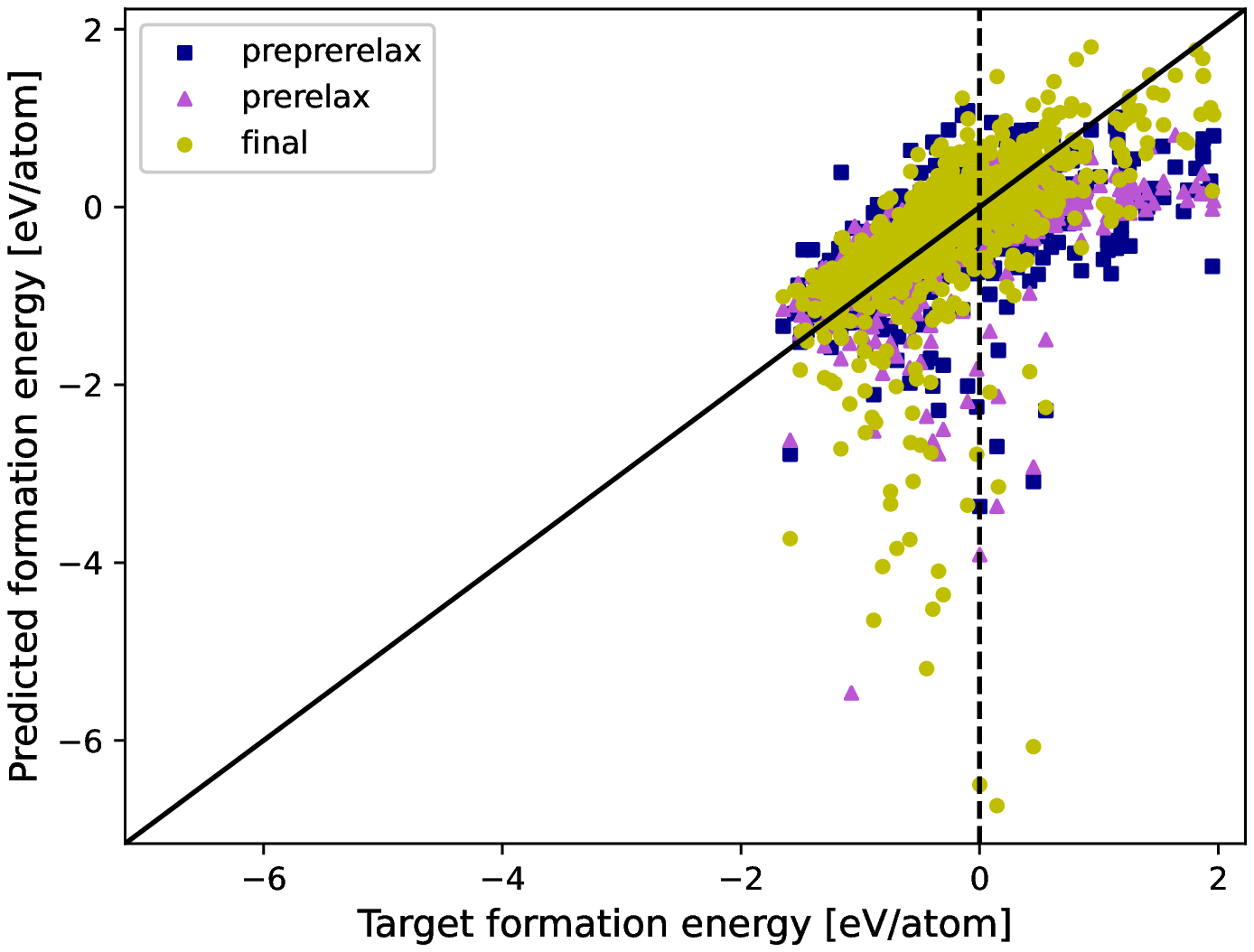}
\caption{\label{fig:from_scratch_pred_vs_actual_hf_60}}
\end{subfigure}
\caption{\label{fig:pred_vs_actual_w_and_wo_finetuning} Predicted vs actual formation energies when predicting formation energies of structures from Hf-Zn-N, using a model (a) pre-trained on Materials Project, before fine-tuning, (b) pre-trained on Materials Project, fine-tuned with 60 datapoints, and (c) trained from scratch using 60 datapoints.}
\end{figure*}

The model pre-trained on Materials Project data has been trained on completely relaxed structures but still, before being fine-tuned, performs worse or similar than guessing 0 eV/atom. This may seem surprising since the model has been trained on a large data set of relaxed structures. It should, however, be kept in mind that the computations for the materials in Materials Project have been performed with different computational parameters than those used for TAATA. To investigate the performance more closely, the predicted formation energies compared to the actual formation energies for the TAATA-Hf data set are shown in Fig.~\ref{fig:xie_pretrained_pred_vs_actual_no_finetuning}. It is clear that without any fine-tuning the model severely underestimates the formation energy for many structures, leading to the high MAE. Fine-tuning with 60 data points (Fig.~\ref{fig:xie_pretrained_pred_vs_actual_60_finetuning}) appears to compensate for this underestimation. We also compare this fine-tuning with training the model from scratch with 60 data points (Fig.~\ref{fig:from_scratch_pred_vs_actual_hf_60}). 

When transferring to the other two data sets, we observe the same qualitative behavior, see the supplemental material \cite{Note2}.

\subsection{Using CGCNN for generating a phase diagram} 
\begin{figure*}
\begin{subfigure}{0.33\linewidth}
\includegraphics[width=\textwidth]{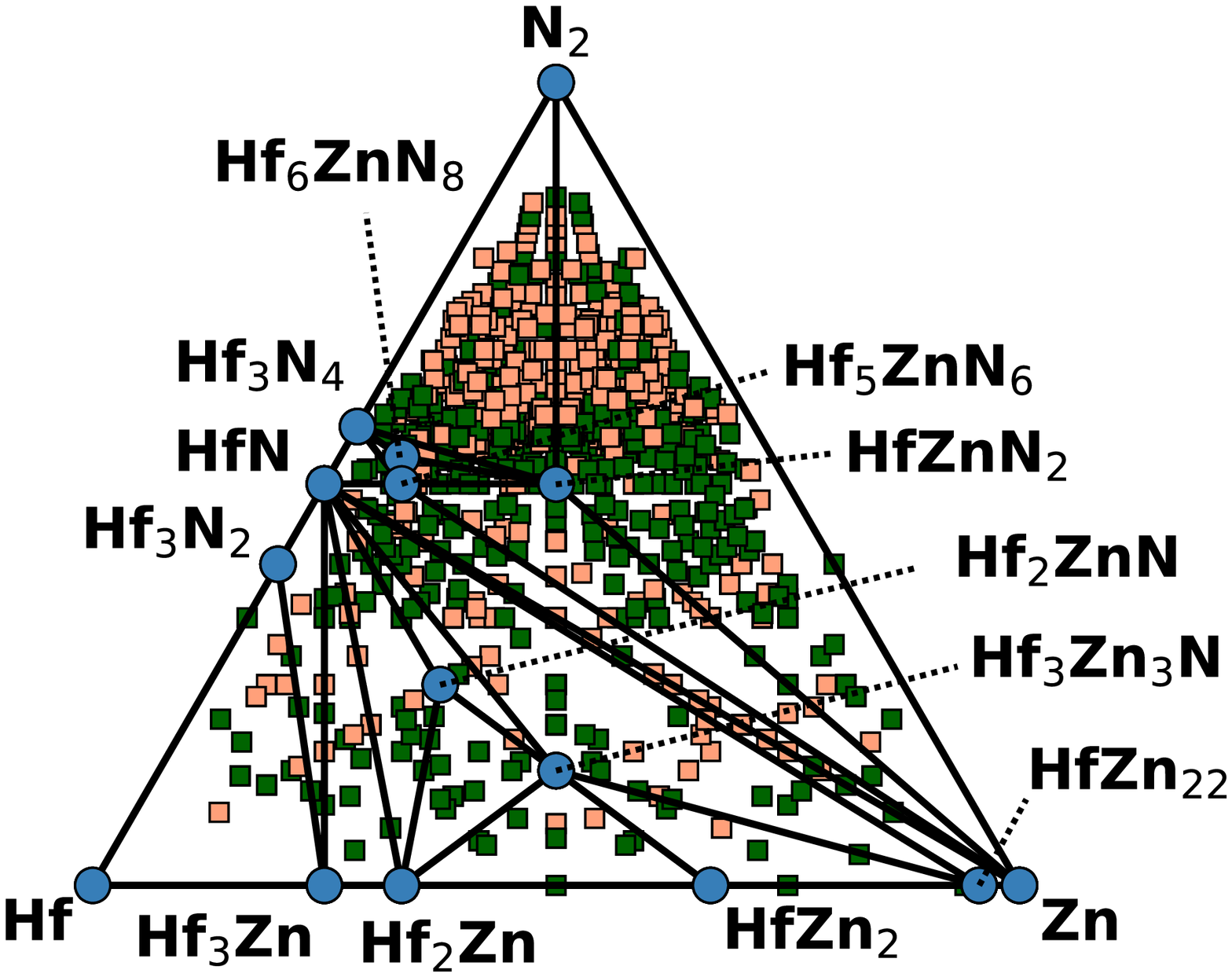}
\caption{\label{fig:pd_prepre}}
\end{subfigure}%
\hfill
\begin{subfigure}{0.33\linewidth}
\includegraphics[width=\textwidth]{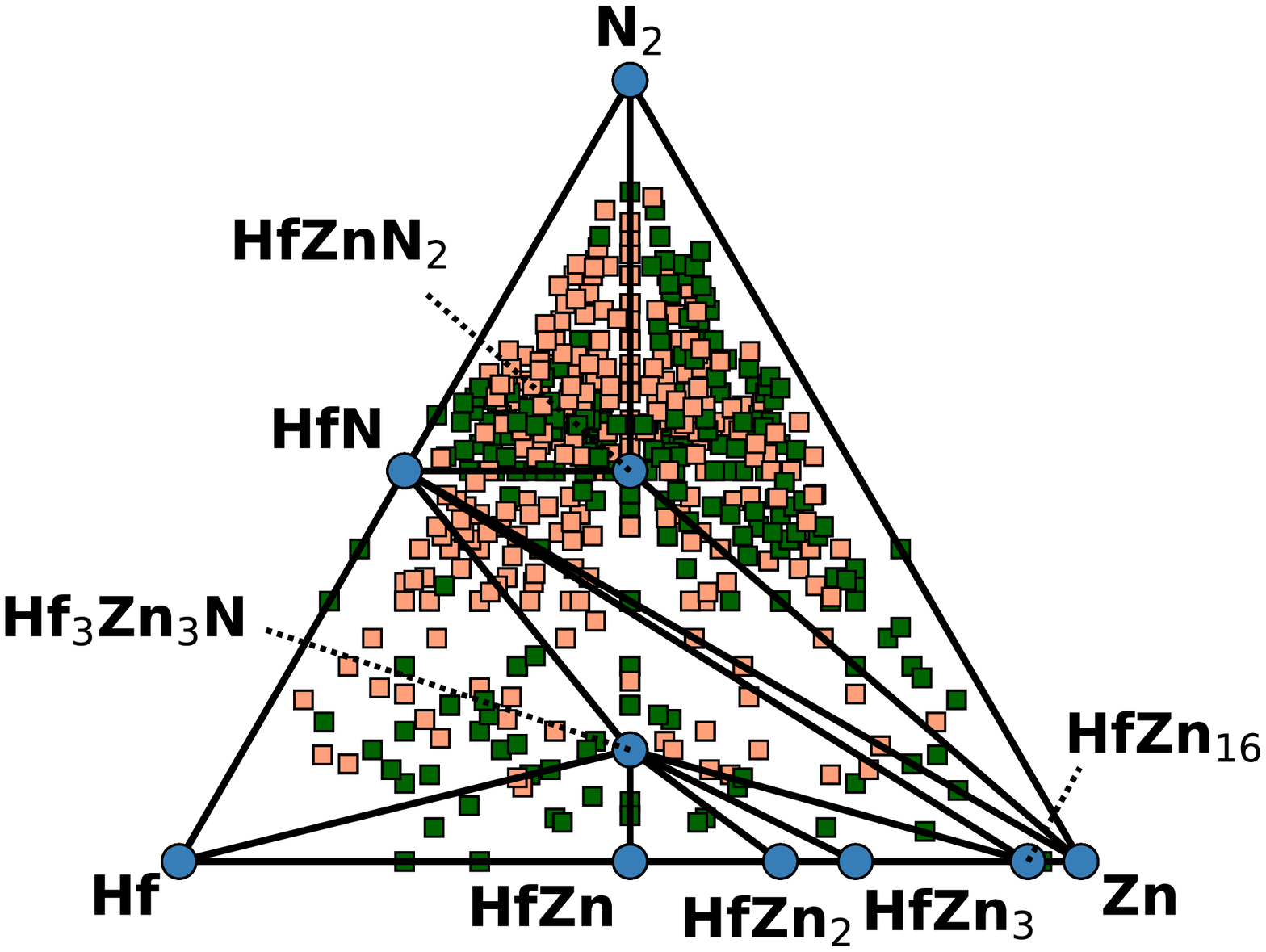}
\caption{\label{fig:pd_pre}}
\end{subfigure}%
\hfill
\begin{subfigure}{0.33\linewidth}
\includegraphics[width=\textwidth]{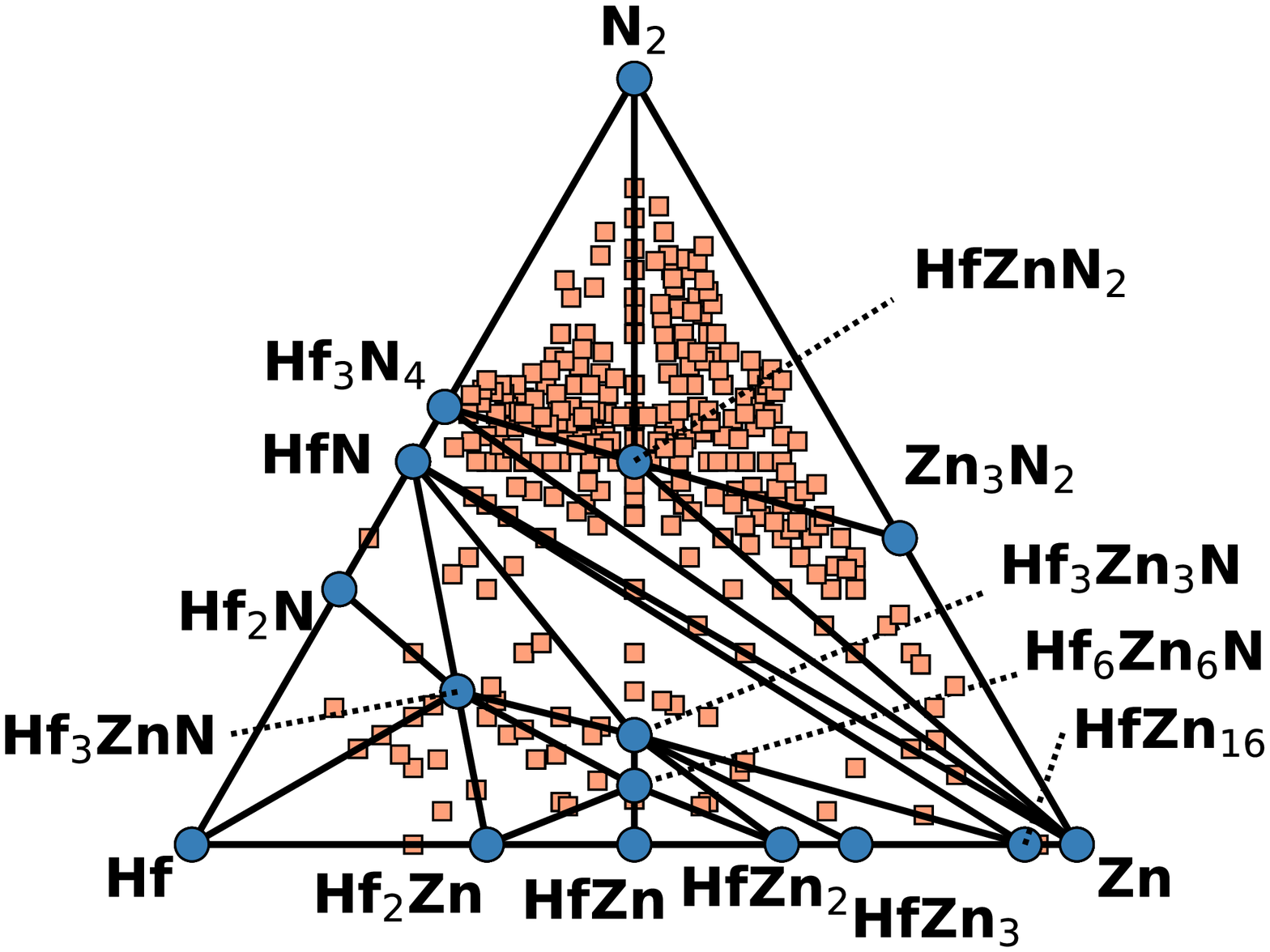}
\caption{\label{fig:pd_final}}
\end{subfigure}
\caption{\label{fig:pd_dev} Development of the predicted phase diagram during structure relaxation after the (a) preprerelax, (b) prerelax and (c) final relaxation level. Blue circles are structures on the convex hull. For the preprerelax and prerelax steps, formation energies are predicted using CGCNN for structures not included in the training set. Dark green squares are structures above the convex hull, but with energy above hull within the threshold $2\sqrt{\sigma^2_a + \sigma_e^2}$ and have therefore been further relaxed. Light red squares are structures above hull with energy above hull outside the threshold, and have thus not been further relaxed. For the final level, all formation energies have been computed with DFT, and red squares indicates structures above the convex hull. The phase diagrams have been plotted with \textit{pymatgen} \cite{ong_python_2013}.}
\end{figure*}

We now demonstrate the scheme described in Sec.~\ref{sec:method} for predicting the full phase diagram of Hf-Zn-N, using a model transferred from Zr-Zn-N by fine-tuning with 60 data points. The development of the phase diagram during structure relaxation is shown in Fig.~\ref{fig:pd_dev}. The final convex hulls of stability is the same as one would arrive to by relaxing and computing formation energies for all candidate structures with DFT. However, by following the suggested scheme,  there is a significant reduction in the number of such calculations that are needed. There are 4779 structures in the initial evaluation set, which are relaxed at the preprerelax step. After predicting the initial phase diagram (Fig.~\ref{fig:pd_prepre}), 2227 structures are kept and go through the prerelax step. From the predicted phase diagram at this level (Fig.~\ref{fig:pd_pre}), 1210 structures are kept and are completely relaxed. Including the training data, 1270 structures are completely relaxed, instead of all of the 4839 candidate structures. Using the average computation times for the different steps, an equivalent of 1417 complete structure relaxations have been performed, which is less than 30 \% of the initial number of candidate structures.

\section{Discussion}
\subsection{Prediction performance on the TAATA data set}%
The results in Fig.~\ref{fig:relax_levels} show that CGCNN performs well when predicting formation energies of the materials in the TAATA data set. The models trained on the largest training sets of relaxed structures show accuracies close to that of the original CGCNN work \cite{xie_crystal_2018}, where a test MAE of 0.039 eV/atom was reported for formation energies of structures in Materials Project. However, the size of the training set for that model was much larger, roughly \thsnd{37} materials. This indicates that this type of data, where materials are confined to only a few elements and for the most part are highly unstable, is indeed manageable for this model. CGCNN also outperforms KRR with the sine-based Coloumb matrix descriptor, further emphasizing that it is a powerful method, and that GNNs in general are an interesting way to further improve the application of machine learning to materials science.

The worsened performance of the model when the materials are not fully relaxed is not surprising, considering that the model relies on information about the geometry. Nevertheless, the model still seems to be able to extract useful information and learn also from these inaccurate descriptions, since the performance is improving as the training set is growing. Additionally, it seems like the error has not saturated at our largest training sets, indicating that the performance could further improve with more data. This can be seen as an indication that the model is able to overcome the fact that the materials are not relaxed and, in some sense, jointly learn the combination of structure relaxation and prediction of the formation energy.

\subsection{Transfer learning}
A model pre-trained on one phase diagram performs very poorly when applied to a different phase diagram without fine-tuning, which is perhaps surprising considering the similarity of the phase diagrams in the TAATA data set. We speculate that the reason for this is that the pre-trained model has only seen three different atom types, and in particular has not seen data with the new atom type introduced in the data set it is transferred to, meaning that it is unable to create a good representation of this element. This suspicion is strengthened by the fact that only a small amount of data for fine-tuning improves the performance considerably.

However, we see the same behavior with poor performance before fine-tuning of the model pre-trained on the Materials Project data set, which has seen a large variety of atoms. In Fig.~\ref{fig:xie_pretrained_pred_vs_actual_no_finetuning} we see that in this case, most of the prediction error stems from the model underestimating the formation energy. We believe that one reason for this is the fact that the model is pre-trained with mostly stable materials, which in general will have lower formation energies. This would then explain the underestimation in the predicted formation energies for TAATA, which mostly contain unstable materials. However, another reason could be the difference in computational parameters used to produce the data, which could lead to systematic errors. Either way, using only a small amount of training data for fine-tuning improves the performance by compensating for the underestimation of formation energies. We therefore conclude that while CGCNN is capable of handling also unstable structures, it is important that it is exposed to this during training. 

Another observation that can be made in Fig. \ref{fig:pred_vs_actual_w_and_wo_finetuning} is that after fine-tuning, the errors seem higher for materials with high formation energies, and that this difference is more prominent for the partially relaxed structures (Fig.~\ref{fig:xie_pretrained_pred_vs_actual_60_finetuning}). This does not seem to be the case when training from scratch (Fig.~\ref{fig:from_scratch_pred_vs_actual_hf_60}), and a reason for this could be that the pre-training on relaxed structures with lower formation energies is lingering even after fine-tuning. One could think of combining a model trained from scratch with a model pre-trained on Materials Project to obtain the best performance for all formation energies. 

The fact that a small amount of extra data gives these significant improvements suggests that, if possible, it can be worthwhile to generate a small annotated training data set for the specific task at hand, instead of relying solely on models trained on some generic database. It seems that when the target data set is a certain phase diagram, pre-training on a data set like Materials Project improves over training a model from scratch, at least when the data availability is small. However, if possible, pre-training on a similar data set to the one under study improves the performance substantially. For example, when predicting formation energies of relaxed structures from TAATA-Hf the test MAE is 0.39 eV/atom when the model was trained from scratch, using 60 data points. When using a model pre-trained on Materials Project and fine-tuning it with 60 datapoints, the error lowers to 0.23 eV/atom. If instead using a model pre-trained on the TAATA-Zr data set and fine-tuned with 60 datapoints, the test MAE is as low as 0.049 eV/atom. Screening a large chemical space of similar compounds is not an uncommon application, and our results show that there is a possibility of very high gains for such applications by using transfer learning.

\subsection{Using CGCNN for generating phase diagrams}
Our method for generating a phase diagram with the help of carefully constructed CGCNN models is an example how our findings can be utilized. We acknowledge that the methodology in this example can be further improved; we have, for example, just chosen a training set at random, and our threshold for when a structure is removed from the relaxation process is chosen more or less arbitrarily. However, we think that the results are promising and indeed demonstrates the potential in using machine learning in combination with DFT in high-throughput screening.

\section{Conclusions}
We have demonstrated that CGCNN can be trained to predict formation energies of materials that belong to a single ternary phase diagram and which for the most part are far from the convex hull. In this scenario, we obtain test MAEs in the range 0.038 to 0.046~eV/atom for three different ternary phase diagrams when using less than \thsnd{4} training data points from the respective phase diagram. We have also shown that CGCNN can make informative predictions also when the structures are not fully relaxed, although with loss of performance compared to fully relaxed structures. Using transfer learning, we are able to transfer a model from one phase diagram to another, using very little additional data; in our case 60 materials are sufficient to get highly accurate predictions. We also compare this with transferring a model pre-trained on a much larger Materials Project data set. Interestingly, a model pre-trained on a different data set performs very poorly, often worse than predicting a constant formation energy for all materials. However, fine tuning the pre-trained model with 60 training data points can improve the performance drastically, especially when the data set used for fine tuning is similar to the target data. From this we also conclude that even though CGCNN can handle unstable materials from a single ternary phase diagram, it needs to be exposed to such materials during training to perform well. Finally, we have illustrated how these insights could be used to aid in the generation of a phase diagram by, during structure relaxation, using carefully constructed CGCNN models to predict the final formation energies and only continue relaxation of the structures that are predicted to lie close to the convex hull of stability. Even though it is a simple approach, it emphasizes that machine learning has a great potential for speeding up high-throughput screening.

\begin{acknowledgments}
RA acknowledges support from the Swedish Research Council (VR) Grant No. 2020-05402  and the Swedish e-Science Centre (SeRC). The computations for the TAATA data set were enabled by resources provided by the Swedish National Infrastructure for Computing (SNIC) at NSC partially funded by the Swedish Research Council through grant agreement no.~2018-05973.

FL acknowledges support from the Swedish Research Council (VR) Grant No. 2020-04122, 
the Swedish Foundation for Strategic Research (SSF) Grant No. ICA16-0015,
the Wallenberg AI, Autonomous Systems and Software Program (WASP) funded by the Knut and Alice Wallenberg Foundation,
and
ELLIIT.
\end{acknowledgments}


\begin{thebibliography}{46}%
\makeatletter
\providecommand \@ifxundefined [1]{%
 \@ifx{#1\undefined}
}%
\providecommand \@ifnum [1]{%
 \ifnum #1\expandafter \@firstoftwo
 \else \expandafter \@secondoftwo
 \fi
}%
\providecommand \@ifx [1]{%
 \ifx #1\expandafter \@firstoftwo
 \else \expandafter \@secondoftwo
 \fi
}%
\providecommand \natexlab [1]{#1}%
\providecommand \enquote  [1]{``#1''}%
\providecommand \bibnamefont  [1]{#1}%
\providecommand \bibfnamefont [1]{#1}%
\providecommand \citenamefont [1]{#1}%
\providecommand \href@noop [0]{\@secondoftwo}%
\providecommand \href [0]{\begingroup \@sanitize@url \@href}%
\providecommand \@href[1]{\@@startlink{#1}\@@href}%
\providecommand \@@href[1]{\endgroup#1\@@endlink}%
\providecommand \@sanitize@url [0]{\catcode `\\12\catcode `\$12\catcode
  `\&12\catcode `\#12\catcode `\^12\catcode `\_12\catcode `\%12\relax}%
\providecommand \@@startlink[1]{}%
\providecommand \@@endlink[0]{}%
\providecommand \url  [0]{\begingroup\@sanitize@url \@url }%
\providecommand \@url [1]{\endgroup\@href {#1}{\urlprefix }}%
\providecommand \urlprefix  [0]{URL }%
\providecommand \Eprint [0]{\href }%
\providecommand \doibase [0]{https://doi.org/}%
\providecommand \selectlanguage [0]{\@gobble}%
\providecommand \bibinfo  [0]{\@secondoftwo}%
\providecommand \bibfield  [0]{\@secondoftwo}%
\providecommand \translation [1]{[#1]}%
\providecommand \BibitemOpen [0]{}%
\providecommand \bibitemStop [0]{}%
\providecommand \bibitemNoStop [0]{.\EOS\space}%
\providecommand \EOS [0]{\spacefactor3000\relax}%
\providecommand \BibitemShut  [1]{\csname bibitem#1\endcsname}%
\let\auto@bib@innerbib\@empty
\bibitem [{\citenamefont {Armiento}(2020)}]{armiento_database-driven_2020}%
  \BibitemOpen
  \bibfield  {author} {\bibinfo {author} {\bibfnamefont {R.}~\bibnamefont
  {Armiento}},\ }\bibfield  {title} {\bibinfo {title} {Database-{{Driven
  High}}-{{Throughput Calculations}} and {{Machine Learning Models}} for
  {{Materials Design}}},\ }in\ \href
  {https://doi.org/10.1007/978-3-030-40245-7_17} {\emph {\bibinfo {booktitle}
  {Machine {{Learning Meets Quantum Physics}}}}},\ \bibinfo {series and number}
  {Lecture {{Notes}} in {{Physics}}},\ \bibinfo {editor} {edited by\ \bibinfo
  {editor} {\bibfnamefont {K.~T.}\ \bibnamefont {Sch{\"u}tt}}, \bibinfo
  {editor} {\bibfnamefont {S.}~\bibnamefont {Chmiela}}, \bibinfo {editor}
  {\bibfnamefont {O.~A.}\ \bibnamefont {{von Lilienfeld}}}, \bibinfo {editor}
  {\bibfnamefont {A.}~\bibnamefont {Tkatchenko}}, \bibinfo {editor}
  {\bibfnamefont {K.}~\bibnamefont {Tsuda}},\ and\ \bibinfo {editor}
  {\bibfnamefont {K.-R.}\ \bibnamefont {M{\"u}ller}}}\ (\bibinfo  {publisher}
  {{Springer International Publishing}},\ \bibinfo {address} {{Cham}},\
  \bibinfo {year} {2020})\ pp.\ \bibinfo {pages} {377--395}\BibitemShut
  {NoStop}%
\bibitem [{\citenamefont {Hohenberg}\ and\ \citenamefont
  {Kohn}(1964)}]{hohenberg_inhomogeneous_1964-1}%
  \BibitemOpen
  \bibfield  {author} {\bibinfo {author} {\bibfnamefont {P.}~\bibnamefont
  {Hohenberg}}\ and\ \bibinfo {author} {\bibfnamefont {W.}~\bibnamefont
  {Kohn}},\ }\bibfield  {title} {\bibinfo {title} {Inhomogeneous {{Electron
  Gas}}},\ }\href {https://doi.org/10.1103/PhysRev.136.B864} {\bibfield
  {journal} {\bibinfo  {journal} {Physical Review}\ }\textbf {\bibinfo {volume}
  {136}},\ \bibinfo {pages} {B864} (\bibinfo {year} {1964})}\BibitemShut
  {NoStop}%
\bibitem [{\citenamefont {Kohn}\ and\ \citenamefont
  {Sham}(1965)}]{kohn_self-consistent_1965-1}%
  \BibitemOpen
  \bibfield  {author} {\bibinfo {author} {\bibfnamefont {W.}~\bibnamefont
  {Kohn}}\ and\ \bibinfo {author} {\bibfnamefont {L.~J.}\ \bibnamefont
  {Sham}},\ }\bibfield  {title} {\bibinfo {title} {Self-{{Consistent Equations
  Including Exchange}} and {{Correlation Effects}}},\ }\href
  {https://doi.org/10.1103/PhysRev.140.A1133} {\bibfield  {journal} {\bibinfo
  {journal} {Physical Review}\ }\textbf {\bibinfo {volume} {140}},\ \bibinfo
  {pages} {A1133} (\bibinfo {year} {1965})}\BibitemShut {NoStop}%
\bibitem [{\citenamefont {Schmidt}\ \emph {et~al.}(2019)\citenamefont
  {Schmidt}, \citenamefont {Marques}, \citenamefont {Botti},\ and\
  \citenamefont {Marques}}]{schmidt_recent_2019}%
  \BibitemOpen
  \bibfield  {author} {\bibinfo {author} {\bibfnamefont {J.}~\bibnamefont
  {Schmidt}}, \bibinfo {author} {\bibfnamefont {M.~R.~G.}\ \bibnamefont
  {Marques}}, \bibinfo {author} {\bibfnamefont {S.}~\bibnamefont {Botti}},\
  and\ \bibinfo {author} {\bibfnamefont {M.~A.~L.}\ \bibnamefont {Marques}},\
  }\bibfield  {title} {\bibinfo {title} {Recent advances and applications of
  machine learning in solid-state materials science},\ }\href
  {https://doi.org/10.1038/s41524-019-0221-0} {\bibfield  {journal} {\bibinfo
  {journal} {npj Computational Materials}\ }\textbf {\bibinfo {volume} {5}},\
  \bibinfo {pages} {1} (\bibinfo {year} {2019})}\BibitemShut {NoStop}%
\bibitem [{\citenamefont {Schleder}\ \emph {et~al.}(2019)\citenamefont
  {Schleder}, \citenamefont {Padilha}, \citenamefont {Acosta}, \citenamefont
  {Costa},\ and\ \citenamefont {Fazzio}}]{schleder_dft_2019}%
  \BibitemOpen
  \bibfield  {author} {\bibinfo {author} {\bibfnamefont {G.~R.}\ \bibnamefont
  {Schleder}}, \bibinfo {author} {\bibfnamefont {A.~C.~M.}\ \bibnamefont
  {Padilha}}, \bibinfo {author} {\bibfnamefont {C.~M.}\ \bibnamefont {Acosta}},
  \bibinfo {author} {\bibfnamefont {M.}~\bibnamefont {Costa}},\ and\ \bibinfo
  {author} {\bibfnamefont {A.}~\bibnamefont {Fazzio}},\ }\bibfield  {title}
  {\bibinfo {title} {From {{DFT}} to machine learning: Recent approaches to
  materials science\textendash a review},\ }\href
  {https://doi.org/10.1088/2515-7639/ab084b} {\bibfield  {journal} {\bibinfo
  {journal} {Journal of Physics: Materials}\ }\textbf {\bibinfo {volume} {2}},\
  \bibinfo {pages} {032001} (\bibinfo {year} {2019})}\BibitemShut {NoStop}%
\bibitem [{\citenamefont {Chen}\ \emph {et~al.}(2020)\citenamefont {Chen},
  \citenamefont {Zuo}, \citenamefont {Ye}, \citenamefont {Li}, \citenamefont
  {Deng},\ and\ \citenamefont {Ong}}]{chen_critical_2020}%
  \BibitemOpen
  \bibfield  {author} {\bibinfo {author} {\bibfnamefont {C.}~\bibnamefont
  {Chen}}, \bibinfo {author} {\bibfnamefont {Y.}~\bibnamefont {Zuo}}, \bibinfo
  {author} {\bibfnamefont {W.}~\bibnamefont {Ye}}, \bibinfo {author}
  {\bibfnamefont {X.}~\bibnamefont {Li}}, \bibinfo {author} {\bibfnamefont
  {Z.}~\bibnamefont {Deng}},\ and\ \bibinfo {author} {\bibfnamefont {S.~P.}\
  \bibnamefont {Ong}},\ }\bibfield  {title} {\bibinfo {title} {A {{Critical
  Review}} of {{Machine Learning}} of {{Energy Materials}}},\ }\href
  {https://doi.org/10.1002/aenm.201903242} {\bibfield  {journal} {\bibinfo
  {journal} {Advanced Energy Materials}\ }\textbf {\bibinfo {volume} {10}},\
  \bibinfo {pages} {1903242} (\bibinfo {year} {2020})}\BibitemShut {NoStop}%
\bibitem [{\citenamefont {Faber}\ \emph {et~al.}(2017)\citenamefont {Faber},
  \citenamefont {Hutchison}, \citenamefont {Huang}, \citenamefont {Gilmer},
  \citenamefont {Schoenholz}, \citenamefont {Dahl}, \citenamefont {Vinyals},
  \citenamefont {Kearnes}, \citenamefont {Riley},\ and\ \citenamefont {{von
  Lilienfeld}}}]{faber_prediction_2017}%
  \BibitemOpen
  \bibfield  {author} {\bibinfo {author} {\bibfnamefont {F.~A.}\ \bibnamefont
  {Faber}}, \bibinfo {author} {\bibfnamefont {L.}~\bibnamefont {Hutchison}},
  \bibinfo {author} {\bibfnamefont {B.}~\bibnamefont {Huang}}, \bibinfo
  {author} {\bibfnamefont {J.}~\bibnamefont {Gilmer}}, \bibinfo {author}
  {\bibfnamefont {S.~S.}\ \bibnamefont {Schoenholz}}, \bibinfo {author}
  {\bibfnamefont {G.~E.}\ \bibnamefont {Dahl}}, \bibinfo {author}
  {\bibfnamefont {O.}~\bibnamefont {Vinyals}}, \bibinfo {author} {\bibfnamefont
  {S.}~\bibnamefont {Kearnes}}, \bibinfo {author} {\bibfnamefont {P.~F.}\
  \bibnamefont {Riley}},\ and\ \bibinfo {author} {\bibfnamefont {O.~A.}\
  \bibnamefont {{von Lilienfeld}}},\ }\bibfield  {title} {\bibinfo {title}
  {Prediction {{Errors}} of {{Molecular Machine Learning Models Lower}} than
  {{Hybrid DFT Error}}},\ }\href {https://doi.org/10.1021/acs.jctc.7b00577}
  {\bibfield  {journal} {\bibinfo  {journal} {Journal of Chemical Theory and
  Computation}\ }\textbf {\bibinfo {volume} {13}},\ \bibinfo {pages} {5255}
  (\bibinfo {year} {2017})}\BibitemShut {NoStop}%
\bibitem [{\citenamefont {Bart{\'o}k}\ \emph {et~al.}(2013)\citenamefont
  {Bart{\'o}k}, \citenamefont {Kondor},\ and\ \citenamefont
  {Cs{\'a}nyi}}]{bartok_representing_2013}%
  \BibitemOpen
  \bibfield  {author} {\bibinfo {author} {\bibfnamefont {A.~P.}\ \bibnamefont
  {Bart{\'o}k}}, \bibinfo {author} {\bibfnamefont {R.}~\bibnamefont {Kondor}},\
  and\ \bibinfo {author} {\bibfnamefont {G.}~\bibnamefont {Cs{\'a}nyi}},\
  }\bibfield  {title} {\bibinfo {title} {On representing chemical
  environments},\ }\href {https://doi.org/10.1103/PhysRevB.87.184115}
  {\bibfield  {journal} {\bibinfo  {journal} {Physical Review B}\ }\textbf
  {\bibinfo {volume} {87}},\ \bibinfo {pages} {184115} (\bibinfo {year}
  {2013})}\BibitemShut {NoStop}%
\bibitem [{\citenamefont {Faber}\ \emph {et~al.}(2015)\citenamefont {Faber},
  \citenamefont {Lindmaa}, \citenamefont {von Lilienfeld},\ and\ \citenamefont
  {Armiento}}]{faber_crystal_2015}%
  \BibitemOpen
  \bibfield  {author} {\bibinfo {author} {\bibfnamefont {F.}~\bibnamefont
  {Faber}}, \bibinfo {author} {\bibfnamefont {A.}~\bibnamefont {Lindmaa}},
  \bibinfo {author} {\bibfnamefont {O.~A.}\ \bibnamefont {von Lilienfeld}},\
  and\ \bibinfo {author} {\bibfnamefont {R.}~\bibnamefont {Armiento}},\
  }\bibfield  {title} {\bibinfo {title} {Crystal structure representations for
  machine learning models of formation energies},\ }\href
  {https://doi.org/10.1002/qua.24917} {\bibfield  {journal} {\bibinfo
  {journal} {International Journal of Quantum Chemistry}\ }\textbf {\bibinfo
  {volume} {115}},\ \bibinfo {pages} {1094} (\bibinfo {year}
  {2015})}\BibitemShut {NoStop}%
\bibitem [{\citenamefont {Faber}\ \emph {et~al.}(2016)\citenamefont {Faber},
  \citenamefont {Lindmaa}, \citenamefont {von Lilienfeld},\ and\ \citenamefont
  {Armiento}}]{faber_machine_2016}%
  \BibitemOpen
  \bibfield  {author} {\bibinfo {author} {\bibfnamefont {F.~A.}\ \bibnamefont
  {Faber}}, \bibinfo {author} {\bibfnamefont {A.}~\bibnamefont {Lindmaa}},
  \bibinfo {author} {\bibfnamefont {O.~A.}\ \bibnamefont {von Lilienfeld}},\
  and\ \bibinfo {author} {\bibfnamefont {R.}~\bibnamefont {Armiento}},\
  }\bibfield  {title} {\bibinfo {title} {Machine learning energies of 2 million
  elpasolite $(ab{C}_{2}{D}_{6})$ crystals},\ }\href
  {https://doi.org/10.1103/PhysRevLett.117.135502} {\bibfield  {journal}
  {\bibinfo  {journal} {Phys. Rev. Lett.}\ }\textbf {\bibinfo {volume} {117}},\
  \bibinfo {pages} {135502} (\bibinfo {year} {2016})}\BibitemShut {NoStop}%
\bibitem [{\citenamefont {Huo}\ and\ \citenamefont
  {Rupp}(2018)}]{huo_unified_2018}%
  \BibitemOpen
  \bibfield  {author} {\bibinfo {author} {\bibfnamefont {H.}~\bibnamefont
  {Huo}}\ and\ \bibinfo {author} {\bibfnamefont {M.}~\bibnamefont {Rupp}},\
  }\bibfield  {title} {\bibinfo {title} {Unified {{Representation}} of
  {{Molecules}} and {{Crystals}} for {{Machine Learning}}},\ }\href@noop {}
  {\bibfield  {journal} {\bibinfo  {journal} {arXiv:1704.06439 [cond-mat,
  physics:physics]}\ } (\bibinfo {year} {2018})},\ \Eprint
  {https://arxiv.org/abs/1704.06439} {arXiv:1704.06439 [cond-mat,
  physics:physics]} \BibitemShut {NoStop}%
\bibitem [{\citenamefont {Szlachta}\ \emph {et~al.}(2014)\citenamefont
  {Szlachta}, \citenamefont {Bart{\'o}k},\ and\ \citenamefont
  {Cs{\'a}nyi}}]{szlachta_accuracy_2014}%
  \BibitemOpen
  \bibfield  {author} {\bibinfo {author} {\bibfnamefont {W.~J.}\ \bibnamefont
  {Szlachta}}, \bibinfo {author} {\bibfnamefont {A.~P.}\ \bibnamefont
  {Bart{\'o}k}},\ and\ \bibinfo {author} {\bibfnamefont {G.}~\bibnamefont
  {Cs{\'a}nyi}},\ }\bibfield  {title} {\bibinfo {title} {Accuracy and
  transferability of {{Gaussian}} approximation potential models for
  tungsten},\ }\href {https://doi.org/10.1103/PhysRevB.90.104108} {\bibfield
  {journal} {\bibinfo  {journal} {Physical Review B}\ }\textbf {\bibinfo
  {volume} {90}},\ \bibinfo {pages} {104108} (\bibinfo {year}
  {2014})}\BibitemShut {NoStop}%
\bibitem [{\citenamefont {Dragoni}\ \emph {et~al.}(2018)\citenamefont
  {Dragoni}, \citenamefont {Daff}, \citenamefont {Cs{\'a}nyi},\ and\
  \citenamefont {Marzari}}]{dragoni_achieving_2018}%
  \BibitemOpen
  \bibfield  {author} {\bibinfo {author} {\bibfnamefont {D.}~\bibnamefont
  {Dragoni}}, \bibinfo {author} {\bibfnamefont {T.~D.}\ \bibnamefont {Daff}},
  \bibinfo {author} {\bibfnamefont {G.}~\bibnamefont {Cs{\'a}nyi}},\ and\
  \bibinfo {author} {\bibfnamefont {N.}~\bibnamefont {Marzari}},\ }\bibfield
  {title} {\bibinfo {title} {Achieving {{DFT}} accuracy with a machine-learning
  interatomic potential: {{Thermomechanics}} and defects in bcc ferromagnetic
  iron},\ }\href {https://doi.org/10.1103/PhysRevMaterials.2.013808} {\bibfield
   {journal} {\bibinfo  {journal} {Physical Review Materials}\ }\textbf
  {\bibinfo {volume} {2}},\ \bibinfo {pages} {013808} (\bibinfo {year}
  {2018})}\BibitemShut {NoStop}%
\bibitem [{\citenamefont {Shapeev}(2016)}]{shapeev_moment_2016}%
  \BibitemOpen
  \bibfield  {author} {\bibinfo {author} {\bibfnamefont {A.~V.}\ \bibnamefont
  {Shapeev}},\ }\bibfield  {title} {\bibinfo {title} {Moment {{Tensor
  Potentials}}: {{A Class}} of {{Systematically Improvable Interatomic
  Potentials}}},\ }\href {https://doi.org/10.1137/15M1054183} {\bibfield
  {journal} {\bibinfo  {journal} {Multiscale Modeling \& Simulation}\ }\textbf
  {\bibinfo {volume} {14}},\ \bibinfo {pages} {1153} (\bibinfo {year}
  {2016})}\BibitemShut {NoStop}%
\bibitem [{\citenamefont {Podryabinkin}\ and\ \citenamefont
  {Shapeev}(2017)}]{podryabinkin_active_2017}%
  \BibitemOpen
  \bibfield  {author} {\bibinfo {author} {\bibfnamefont {E.~V.}\ \bibnamefont
  {Podryabinkin}}\ and\ \bibinfo {author} {\bibfnamefont {A.~V.}\ \bibnamefont
  {Shapeev}},\ }\bibfield  {title} {\bibinfo {title} {Active learning of
  linearly parametrized interatomic potentials},\ }\href
  {https://doi.org/10.1016/j.commatsci.2017.08.031} {\bibfield  {journal}
  {\bibinfo  {journal} {Computational Materials Science}\ }\textbf {\bibinfo
  {volume} {140}},\ \bibinfo {pages} {171} (\bibinfo {year}
  {2017})}\BibitemShut {NoStop}%
\bibitem [{\citenamefont {Gubaev}\ \emph {et~al.}(2019)\citenamefont {Gubaev},
  \citenamefont {Podryabinkin}, \citenamefont {Hart},\ and\ \citenamefont
  {Shapeev}}]{gubaev_accelerating_2019}%
  \BibitemOpen
  \bibfield  {author} {\bibinfo {author} {\bibfnamefont {K.}~\bibnamefont
  {Gubaev}}, \bibinfo {author} {\bibfnamefont {E.~V.}\ \bibnamefont
  {Podryabinkin}}, \bibinfo {author} {\bibfnamefont {G.~L.~W.}\ \bibnamefont
  {Hart}},\ and\ \bibinfo {author} {\bibfnamefont {A.~V.}\ \bibnamefont
  {Shapeev}},\ }\bibfield  {title} {\bibinfo {title} {Accelerating
  high-throughput searches for new alloys with active learning of interatomic
  potentials},\ }\href {https://doi.org/10.1016/j.commatsci.2018.09.031}
  {\bibfield  {journal} {\bibinfo  {journal} {Computational Materials Science}\
  }\textbf {\bibinfo {volume} {156}},\ \bibinfo {pages} {148} (\bibinfo {year}
  {2019})}\BibitemShut {NoStop}%
\bibitem [{\citenamefont {Duvenaud}\ \emph {et~al.}(2015)\citenamefont
  {Duvenaud}, \citenamefont {Maclaurin}, \citenamefont {Iparraguirre},
  \citenamefont {Bombarell}, \citenamefont {Hirzel}, \citenamefont
  {{Aspuru-Guzik}},\ and\ \citenamefont {Adams}}]{duvenaud_convolutional_2015}%
  \BibitemOpen
  \bibfield  {author} {\bibinfo {author} {\bibfnamefont {D.~K.}\ \bibnamefont
  {Duvenaud}}, \bibinfo {author} {\bibfnamefont {D.}~\bibnamefont {Maclaurin}},
  \bibinfo {author} {\bibfnamefont {J.}~\bibnamefont {Iparraguirre}}, \bibinfo
  {author} {\bibfnamefont {R.}~\bibnamefont {Bombarell}}, \bibinfo {author}
  {\bibfnamefont {T.}~\bibnamefont {Hirzel}}, \bibinfo {author} {\bibfnamefont
  {A.}~\bibnamefont {{Aspuru-Guzik}}},\ and\ \bibinfo {author} {\bibfnamefont
  {R.~P.}\ \bibnamefont {Adams}},\ }\bibfield  {title} {\bibinfo {title}
  {Convolutional {{Networks}} on {{Graphs}} for {{Learning Molecular
  Fingerprints}}},\ }\href@noop {} {\bibfield  {journal} {\bibinfo  {journal}
  {Advances in Neural Information Processing Systems}\ }\textbf {\bibinfo
  {volume} {28}},\ \bibinfo {pages} {2224} (\bibinfo {year}
  {2015})}\BibitemShut {NoStop}%
\bibitem [{\citenamefont {Kearnes}\ \emph {et~al.}(2016)\citenamefont
  {Kearnes}, \citenamefont {McCloskey}, \citenamefont {Berndl}, \citenamefont
  {Pande},\ and\ \citenamefont {Riley}}]{kearnes_molecular_2016}%
  \BibitemOpen
  \bibfield  {author} {\bibinfo {author} {\bibfnamefont {S.}~\bibnamefont
  {Kearnes}}, \bibinfo {author} {\bibfnamefont {K.}~\bibnamefont {McCloskey}},
  \bibinfo {author} {\bibfnamefont {M.}~\bibnamefont {Berndl}}, \bibinfo
  {author} {\bibfnamefont {V.}~\bibnamefont {Pande}},\ and\ \bibinfo {author}
  {\bibfnamefont {P.}~\bibnamefont {Riley}},\ }\bibfield  {title} {\bibinfo
  {title} {Molecular graph convolutions: Moving beyond fingerprints},\ }\href
  {https://doi.org/10.1007/s10822-016-9938-8} {\bibfield  {journal} {\bibinfo
  {journal} {Journal of Computer-Aided Molecular Design}\ }\textbf {\bibinfo
  {volume} {30}},\ \bibinfo {pages} {595} (\bibinfo {year} {2016})}\BibitemShut
  {NoStop}%
\bibitem [{\citenamefont {Xie}\ and\ \citenamefont
  {Grossman}(2018)}]{xie_crystal_2018}%
  \BibitemOpen
  \bibfield  {author} {\bibinfo {author} {\bibfnamefont {T.}~\bibnamefont
  {Xie}}\ and\ \bibinfo {author} {\bibfnamefont {J.~C.}\ \bibnamefont
  {Grossman}},\ }\bibfield  {title} {\bibinfo {title} {Crystal {{Graph
  Convolutional Neural Networks}} for an {{Accurate}} and {{Interpretable
  Prediction}} of {{Material Properties}}},\ }\href
  {https://doi.org/10.1103/PhysRevLett.120.145301} {\bibfield  {journal}
  {\bibinfo  {journal} {Physical Review Letters}\ }\textbf {\bibinfo {volume}
  {120}},\ \bibinfo {pages} {145301} (\bibinfo {year} {2018})}\BibitemShut
  {NoStop}%
\bibitem [{\citenamefont {Jain}\ \emph {et~al.}(2013)\citenamefont {Jain},
  \citenamefont {Ong}, \citenamefont {Hautier}, \citenamefont {Chen},
  \citenamefont {Richards}, \citenamefont {Dacek}, \citenamefont {Cholia},
  \citenamefont {Gunter}, \citenamefont {Skinner}, \citenamefont {Ceder},\ and\
  \citenamefont {a.~Persson}}]{Jain2013}%
  \BibitemOpen
  \bibfield  {author} {\bibinfo {author} {\bibfnamefont {A.}~\bibnamefont
  {Jain}}, \bibinfo {author} {\bibfnamefont {S.~P.}\ \bibnamefont {Ong}},
  \bibinfo {author} {\bibfnamefont {G.}~\bibnamefont {Hautier}}, \bibinfo
  {author} {\bibfnamefont {W.}~\bibnamefont {Chen}}, \bibinfo {author}
  {\bibfnamefont {W.~D.}\ \bibnamefont {Richards}}, \bibinfo {author}
  {\bibfnamefont {S.}~\bibnamefont {Dacek}}, \bibinfo {author} {\bibfnamefont
  {S.}~\bibnamefont {Cholia}}, \bibinfo {author} {\bibfnamefont
  {D.}~\bibnamefont {Gunter}}, \bibinfo {author} {\bibfnamefont
  {D.}~\bibnamefont {Skinner}}, \bibinfo {author} {\bibfnamefont
  {G.}~\bibnamefont {Ceder}},\ and\ \bibinfo {author} {\bibfnamefont
  {K.}~\bibnamefont {a.~Persson}},\ }\bibfield  {title} {\bibinfo {title} {The
  {{Materials Project}}: {{A}} materials genome approach to accelerating
  materials innovation},\ }\href {https://doi.org/10.1063/1.4812323} {\bibfield
   {journal} {\bibinfo  {journal} {APL Materials}\ }\textbf {\bibinfo {volume}
  {1}},\ \bibinfo {pages} {011002} (\bibinfo {year} {2013})}\BibitemShut
  {NoStop}%
\bibitem [{\citenamefont {Sanyal}\ \emph {et~al.}(2018)\citenamefont {Sanyal},
  \citenamefont {Balachandran}, \citenamefont {Yadati}, \citenamefont {Kumar},
  \citenamefont {Rajagopalan}, \citenamefont {Sanyal},\ and\ \citenamefont
  {Talukdar}}]{sanyal_mt-cgcnn_2018}%
  \BibitemOpen
  \bibfield  {author} {\bibinfo {author} {\bibfnamefont {S.}~\bibnamefont
  {Sanyal}}, \bibinfo {author} {\bibfnamefont {J.}~\bibnamefont
  {Balachandran}}, \bibinfo {author} {\bibfnamefont {N.}~\bibnamefont
  {Yadati}}, \bibinfo {author} {\bibfnamefont {A.}~\bibnamefont {Kumar}},
  \bibinfo {author} {\bibfnamefont {P.}~\bibnamefont {Rajagopalan}}, \bibinfo
  {author} {\bibfnamefont {S.}~\bibnamefont {Sanyal}},\ and\ \bibinfo {author}
  {\bibfnamefont {P.}~\bibnamefont {Talukdar}},\ }\bibfield  {title} {\bibinfo
  {title} {{{MT}}-{{CGCNN}}: {{Integrating Crystal Graph Convolutional Neural
  Network}} with {{Multitask Learning}} for {{Material Property Prediction}}},\
  }\href@noop {} {\bibfield  {journal} {\bibinfo  {journal} {arXiv:1811.05660
  [cond-mat, stat]}\ } (\bibinfo {year} {2018})},\ \Eprint
  {https://arxiv.org/abs/1811.05660} {arXiv:1811.05660 [cond-mat, stat]}
  \BibitemShut {NoStop}%
\bibitem [{\citenamefont {Chen}\ \emph {et~al.}(2019)\citenamefont {Chen},
  \citenamefont {Ye}, \citenamefont {Zuo}, \citenamefont {Zheng},\ and\
  \citenamefont {Ong}}]{chen_graph_2019}%
  \BibitemOpen
  \bibfield  {author} {\bibinfo {author} {\bibfnamefont {C.}~\bibnamefont
  {Chen}}, \bibinfo {author} {\bibfnamefont {W.}~\bibnamefont {Ye}}, \bibinfo
  {author} {\bibfnamefont {Y.}~\bibnamefont {Zuo}}, \bibinfo {author}
  {\bibfnamefont {C.}~\bibnamefont {Zheng}},\ and\ \bibinfo {author}
  {\bibfnamefont {S.~P.}\ \bibnamefont {Ong}},\ }\bibfield  {title} {\bibinfo
  {title} {Graph {{Networks}} as a {{Universal Machine Learning Framework}} for
  {{Molecules}} and {{Crystals}}},\ }\href
  {https://doi.org/10.1021/acs.chemmater.9b01294} {\bibfield  {journal}
  {\bibinfo  {journal} {Chemistry of Materials}\ }\textbf {\bibinfo {volume}
  {31}},\ \bibinfo {pages} {3564} (\bibinfo {year} {2019})}\BibitemShut
  {NoStop}%
\bibitem [{\citenamefont {Park}\ and\ \citenamefont
  {Wolverton}(2020)}]{park_developing_2020}%
  \BibitemOpen
  \bibfield  {author} {\bibinfo {author} {\bibfnamefont {C.~W.}\ \bibnamefont
  {Park}}\ and\ \bibinfo {author} {\bibfnamefont {C.}~\bibnamefont
  {Wolverton}},\ }\bibfield  {title} {\bibinfo {title} {Developing an improved
  crystal graph convolutional neural network framework for accelerated
  materials discovery},\ }\href
  {https://doi.org/10.1103/PhysRevMaterials.4.063801} {\bibfield  {journal}
  {\bibinfo  {journal} {Physical Review Materials}\ }\textbf {\bibinfo {volume}
  {4}},\ \bibinfo {pages} {063801} (\bibinfo {year} {2020})}\BibitemShut
  {NoStop}%
\bibitem [{\citenamefont {Bergerhoff}\ \emph {et~al.}(1983)\citenamefont
  {Bergerhoff}, \citenamefont {Hundt}, \citenamefont {Sievers},\ and\
  \citenamefont {Brown}}]{bergerhoff_inorganic_1983-1}%
  \BibitemOpen
  \bibfield  {author} {\bibinfo {author} {\bibfnamefont {G.}~\bibnamefont
  {Bergerhoff}}, \bibinfo {author} {\bibfnamefont {R.}~\bibnamefont {Hundt}},
  \bibinfo {author} {\bibfnamefont {R.}~\bibnamefont {Sievers}},\ and\ \bibinfo
  {author} {\bibfnamefont {I.~D.}\ \bibnamefont {Brown}},\ }\bibfield  {title}
  {\bibinfo {title} {The inorganic crystal structure data base},\ }\href
  {https://doi.org/10.1021/ci00038a003} {\bibfield  {journal} {\bibinfo
  {journal} {Journal of Chemical Information and Computer Sciences}\ }\textbf
  {\bibinfo {volume} {23}},\ \bibinfo {pages} {66} (\bibinfo {year}
  {1983})}\BibitemShut {NoStop}%
\bibitem [{\citenamefont {Downs}\ and\ \citenamefont
  {{Hall-Wallace}}(2003)}]{Downs2003}%
  \BibitemOpen
  \bibfield  {author} {\bibinfo {author} {\bibfnamefont {R.~T.}\ \bibnamefont
  {Downs}}\ and\ \bibinfo {author} {\bibfnamefont {M.}~\bibnamefont
  {{Hall-Wallace}}},\ }\bibfield  {title} {\bibinfo {title} {The american
  mineralogist crystal structure database},\ }\href@noop {} {\bibfield
  {journal} {\bibinfo  {journal} {American Mineralogist}\ }\textbf {\bibinfo
  {volume} {88}},\ \bibinfo {pages} {247} (\bibinfo {year} {2003})}\BibitemShut
  {NoStop}%
\bibitem [{\citenamefont {Gra{\v z}ulis}\ \emph {et~al.}(2009)\citenamefont
  {Gra{\v z}ulis}, \citenamefont {Chateigner}, \citenamefont {Downs},
  \citenamefont {Yokochi}, \citenamefont {Quir{\'o}s}, \citenamefont
  {Lutterotti}, \citenamefont {Manakova}, \citenamefont {Butkus}, \citenamefont
  {Moeck},\ and\ \citenamefont {Le~Bail}}]{grazulis_crystallography_2009}%
  \BibitemOpen
  \bibfield  {author} {\bibinfo {author} {\bibfnamefont {S.}~\bibnamefont
  {Gra{\v z}ulis}}, \bibinfo {author} {\bibfnamefont {D.}~\bibnamefont
  {Chateigner}}, \bibinfo {author} {\bibfnamefont {R.~T.}\ \bibnamefont
  {Downs}}, \bibinfo {author} {\bibfnamefont {A.~F.~T.}\ \bibnamefont
  {Yokochi}}, \bibinfo {author} {\bibfnamefont {M.}~\bibnamefont {Quir{\'o}s}},
  \bibinfo {author} {\bibfnamefont {L.}~\bibnamefont {Lutterotti}}, \bibinfo
  {author} {\bibfnamefont {E.}~\bibnamefont {Manakova}}, \bibinfo {author}
  {\bibfnamefont {J.}~\bibnamefont {Butkus}}, \bibinfo {author} {\bibfnamefont
  {P.}~\bibnamefont {Moeck}},\ and\ \bibinfo {author} {\bibfnamefont
  {A.}~\bibnamefont {Le~Bail}},\ }\bibfield  {title} {\bibinfo {title}
  {Crystallography {{Open Database}} \textendash{} an open-access collection of
  crystal structures},\ }\href {https://doi.org/10.1107/S0021889809016690}
  {\bibfield  {journal} {\bibinfo  {journal} {Journal of Applied
  Crystallography}\ }\textbf {\bibinfo {volume} {42}},\ \bibinfo {pages} {726}
  (\bibinfo {year} {2009})}\BibitemShut {NoStop}%
\bibitem [{\citenamefont {Gra{\v z}ulis}\ \emph {et~al.}(2012)\citenamefont
  {Gra{\v z}ulis}, \citenamefont {Da{\v s}kevi{\v c}}, \citenamefont {Merkys},
  \citenamefont {Chateigner}, \citenamefont {Lutterotti}, \citenamefont
  {Quir{\'o}s}, \citenamefont {Serebryanaya}, \citenamefont {Moeck},
  \citenamefont {Downs},\ and\ \citenamefont
  {Le~Bail}}]{grazulis_crystallography_2012}%
  \BibitemOpen
  \bibfield  {author} {\bibinfo {author} {\bibfnamefont {S.}~\bibnamefont
  {Gra{\v z}ulis}}, \bibinfo {author} {\bibfnamefont {A.}~\bibnamefont {Da{\v
  s}kevi{\v c}}}, \bibinfo {author} {\bibfnamefont {A.}~\bibnamefont {Merkys}},
  \bibinfo {author} {\bibfnamefont {D.}~\bibnamefont {Chateigner}}, \bibinfo
  {author} {\bibfnamefont {L.}~\bibnamefont {Lutterotti}}, \bibinfo {author}
  {\bibfnamefont {M.}~\bibnamefont {Quir{\'o}s}}, \bibinfo {author}
  {\bibfnamefont {N.~R.}\ \bibnamefont {Serebryanaya}}, \bibinfo {author}
  {\bibfnamefont {P.}~\bibnamefont {Moeck}}, \bibinfo {author} {\bibfnamefont
  {R.~T.}\ \bibnamefont {Downs}},\ and\ \bibinfo {author} {\bibfnamefont
  {A.}~\bibnamefont {Le~Bail}},\ }\bibfield  {title} {\bibinfo {title}
  {Crystallography {{Open Database}} ({{COD}}): An open-access collection of
  crystal structures and platform for world-wide collaboration},\ }\href
  {https://doi.org/10.1093/nar/gkr900} {\bibfield  {journal} {\bibinfo
  {journal} {Nucleic Acids Research}\ }\textbf {\bibinfo {volume} {40}},\
  \bibinfo {pages} {D420} (\bibinfo {year} {2012})}\BibitemShut {NoStop}%
\bibitem [{\citenamefont {Noh}\ \emph {et~al.}(2020)\citenamefont {Noh},
  \citenamefont {Gu}, \citenamefont {Kim},\ and\ \citenamefont
  {Jung}}]{noh_uncertainty-quantified_2020}%
  \BibitemOpen
  \bibfield  {author} {\bibinfo {author} {\bibfnamefont {J.}~\bibnamefont
  {Noh}}, \bibinfo {author} {\bibfnamefont {G.~H.}\ \bibnamefont {Gu}},
  \bibinfo {author} {\bibfnamefont {S.}~\bibnamefont {Kim}},\ and\ \bibinfo
  {author} {\bibfnamefont {Y.}~\bibnamefont {Jung}},\ }\bibfield  {title}
  {\bibinfo {title} {Uncertainty-{{Quantified Hybrid Machine
  Learning}}/{{Density Functional Theory High Throughput Screening Method}} for
  {{Crystals}}},\ }\href {https://doi.org/10.1021/acs.jcim.0c00003} {\bibfield
  {journal} {\bibinfo  {journal} {Journal of Chemical Information and
  Modeling}\ }\textbf {\bibinfo {volume} {60}},\ \bibinfo {pages} {1996}
  (\bibinfo {year} {2020})}\BibitemShut {NoStop}%
\bibitem [{\citenamefont {Tholander}\ \emph {et~al.}(2016)\citenamefont
  {Tholander}, \citenamefont {Andersson}, \citenamefont {Armiento},
  \citenamefont {Tasn{\'a}di},\ and\ \citenamefont
  {Alling}}]{tholander_strong_2016}%
  \BibitemOpen
  \bibfield  {author} {\bibinfo {author} {\bibfnamefont {C.}~\bibnamefont
  {Tholander}}, \bibinfo {author} {\bibfnamefont {C.~B.~A.}\ \bibnamefont
  {Andersson}}, \bibinfo {author} {\bibfnamefont {R.}~\bibnamefont {Armiento}},
  \bibinfo {author} {\bibfnamefont {F.}~\bibnamefont {Tasn{\'a}di}},\ and\
  \bibinfo {author} {\bibfnamefont {B.}~\bibnamefont {Alling}},\ }\bibfield
  {title} {\bibinfo {title} {Strong piezoelectric response in stable
  {{TiZnN2}}, {{ZrZnN2}}, and {{HfZnN2}} found by ab initio high-throughput
  approach},\ }\href {https://doi.org/10.1063/1.4971248} {\bibfield  {journal}
  {\bibinfo  {journal} {Journal of Applied Physics}\ }\textbf {\bibinfo
  {volume} {120}},\ \bibinfo {pages} {225102} (\bibinfo {year}
  {2016})}\BibitemShut {NoStop}%
\bibitem [{\citenamefont {Ong}\ \emph {et~al.}(2008)\citenamefont {Ong},
  \citenamefont {Wang}, \citenamefont {Kang},\ and\ \citenamefont
  {Ceder}}]{ong_lifepo2_2008}%
  \BibitemOpen
  \bibfield  {author} {\bibinfo {author} {\bibfnamefont {S.~P.}\ \bibnamefont
  {Ong}}, \bibinfo {author} {\bibfnamefont {L.}~\bibnamefont {Wang}}, \bibinfo
  {author} {\bibfnamefont {B.}~\bibnamefont {Kang}},\ and\ \bibinfo {author}
  {\bibfnamefont {G.}~\bibnamefont {Ceder}},\ }\bibfield  {title} {\bibinfo
  {title} {{{Li}}-{{Fe}}-{{P}}-{{O2 Phase Diagram}} from {{First Principles
  Calculations}}},\ }\href {https://doi.org/10.1021/cm702327g} {\bibfield
  {journal} {\bibinfo  {journal} {Chemistry of Materials}\ }\textbf {\bibinfo
  {volume} {20}},\ \bibinfo {pages} {1798} (\bibinfo {year}
  {2008})}\BibitemShut {NoStop}%
\bibitem [{\citenamefont {Ong}\ \emph {et~al.}(2010)\citenamefont {Ong},
  \citenamefont {Jain}, \citenamefont {Hautier}, \citenamefont {Kang},\ and\
  \citenamefont {Ceder}}]{ong_thermal_2010}%
  \BibitemOpen
  \bibfield  {author} {\bibinfo {author} {\bibfnamefont {S.~P.}\ \bibnamefont
  {Ong}}, \bibinfo {author} {\bibfnamefont {A.}~\bibnamefont {Jain}}, \bibinfo
  {author} {\bibfnamefont {G.}~\bibnamefont {Hautier}}, \bibinfo {author}
  {\bibfnamefont {B.}~\bibnamefont {Kang}},\ and\ \bibinfo {author}
  {\bibfnamefont {G.}~\bibnamefont {Ceder}},\ }\bibfield  {title} {\bibinfo
  {title} {Thermal stabilities of delithiated olivine {{MPO4}} ({{M}}={{Fe}},
  {{Mn}}) cathodes investigated using first principles calculations},\ }\href
  {https://doi.org/10.1016/j.elecom.2010.01.010} {\bibfield  {journal}
  {\bibinfo  {journal} {Electrochemistry Communications}\ }\textbf {\bibinfo
  {volume} {12}},\ \bibinfo {pages} {427} (\bibinfo {year} {2010})}\BibitemShut
  {NoStop}%
\bibitem [{\citenamefont {Kirklin}\ \emph {et~al.}(2015)\citenamefont
  {Kirklin}, \citenamefont {Saal}, \citenamefont {Meredig}, \citenamefont
  {Thompson}, \citenamefont {Doak}, \citenamefont {Aykol}, \citenamefont
  {R{\"u}hl},\ and\ \citenamefont {Wolverton}}]{kirklin_open_2015}%
  \BibitemOpen
  \bibfield  {author} {\bibinfo {author} {\bibfnamefont {S.}~\bibnamefont
  {Kirklin}}, \bibinfo {author} {\bibfnamefont {J.~E.}\ \bibnamefont {Saal}},
  \bibinfo {author} {\bibfnamefont {B.}~\bibnamefont {Meredig}}, \bibinfo
  {author} {\bibfnamefont {A.}~\bibnamefont {Thompson}}, \bibinfo {author}
  {\bibfnamefont {J.~W.}\ \bibnamefont {Doak}}, \bibinfo {author}
  {\bibfnamefont {M.}~\bibnamefont {Aykol}}, \bibinfo {author} {\bibfnamefont
  {S.}~\bibnamefont {R{\"u}hl}},\ and\ \bibinfo {author} {\bibfnamefont
  {C.}~\bibnamefont {Wolverton}},\ }\bibfield  {title} {\bibinfo {title} {The
  {{Open Quantum Materials Database}} ({{OQMD}}): Assessing the accuracy of
  {{DFT}} formation energies},\ }\href
  {https://doi.org/10.1038/npjcompumats.2015.10} {\bibfield  {journal}
  {\bibinfo  {journal} {npj Computational Materials}\ }\textbf {\bibinfo
  {volume} {1}},\ \bibinfo {pages} {1} (\bibinfo {year} {2015})}\BibitemShut
  {NoStop}%
\bibitem [{\citenamefont {Hautier}\ \emph {et~al.}(2011)\citenamefont
  {Hautier}, \citenamefont {Fischer}, \citenamefont {Ehrlacher}, \citenamefont
  {Jain},\ and\ \citenamefont {Ceder}}]{hautier_data_2011}%
  \BibitemOpen
  \bibfield  {author} {\bibinfo {author} {\bibfnamefont {G.}~\bibnamefont
  {Hautier}}, \bibinfo {author} {\bibfnamefont {C.}~\bibnamefont {Fischer}},
  \bibinfo {author} {\bibfnamefont {V.}~\bibnamefont {Ehrlacher}}, \bibinfo
  {author} {\bibfnamefont {A.}~\bibnamefont {Jain}},\ and\ \bibinfo {author}
  {\bibfnamefont {G.}~\bibnamefont {Ceder}},\ }\bibfield  {title} {\bibinfo
  {title} {Data {{Mined Ionic Substitutions}} for the {{Discovery}} of {{New
  Compounds}}},\ }\href {https://doi.org/10.1021/ic102031h} {\bibfield
  {journal} {\bibinfo  {journal} {Inorganic Chemistry}\ }\textbf {\bibinfo
  {volume} {50}},\ \bibinfo {pages} {656} (\bibinfo {year} {2011})}\BibitemShut
  {NoStop}%
\bibitem [{\citenamefont {Gilmer}\ \emph {et~al.}(2017)\citenamefont {Gilmer},
  \citenamefont {Schoenholz}, \citenamefont {Riley}, \citenamefont {Vinyals},\
  and\ \citenamefont {Dahl}}]{gilmer_neural_2017}%
  \BibitemOpen
  \bibfield  {author} {\bibinfo {author} {\bibfnamefont {J.}~\bibnamefont
  {Gilmer}}, \bibinfo {author} {\bibfnamefont {S.~S.}\ \bibnamefont
  {Schoenholz}}, \bibinfo {author} {\bibfnamefont {P.~F.}\ \bibnamefont
  {Riley}}, \bibinfo {author} {\bibfnamefont {O.}~\bibnamefont {Vinyals}},\
  and\ \bibinfo {author} {\bibfnamefont {G.~E.}\ \bibnamefont {Dahl}},\
  }\bibfield  {title} {\bibinfo {title} {Neural {{Message Passing}} for
  {{Quantum Chemistry}}},\ }in\ \href@noop {} {\emph {\bibinfo {booktitle}
  {International {{Conference}} on {{Machine Learning}}}}}\ (\bibinfo
  {publisher} {{PMLR}},\ \bibinfo {year} {2017})\ pp.\ \bibinfo {pages}
  {1263--1272}\BibitemShut {NoStop}%
\bibitem [{\citenamefont {Sch{\"u}tt}\ \emph {et~al.}(2018)\citenamefont
  {Sch{\"u}tt}, \citenamefont {Sauceda}, \citenamefont {Kindermans},
  \citenamefont {Tkatchenko},\ and\ \citenamefont
  {M{\"u}ller}}]{schutt_schnet_2018}%
  \BibitemOpen
  \bibfield  {author} {\bibinfo {author} {\bibfnamefont {K.~T.}\ \bibnamefont
  {Sch{\"u}tt}}, \bibinfo {author} {\bibfnamefont {H.~E.}\ \bibnamefont
  {Sauceda}}, \bibinfo {author} {\bibfnamefont {P.-J.}\ \bibnamefont
  {Kindermans}}, \bibinfo {author} {\bibfnamefont {A.}~\bibnamefont
  {Tkatchenko}},\ and\ \bibinfo {author} {\bibfnamefont {K.-R.}\ \bibnamefont
  {M{\"u}ller}},\ }\bibfield  {title} {\bibinfo {title} {{{SchNet}}
  \textendash{} {{A}} deep learning architecture for molecules and materials},\
  }\href {https://doi.org/10.1063/1.5019779} {\bibfield  {journal} {\bibinfo
  {journal} {The Journal of Chemical Physics}\ }\textbf {\bibinfo {volume}
  {148}},\ \bibinfo {pages} {241722} (\bibinfo {year} {2018})}\BibitemShut
  {NoStop}%
\bibitem [{\citenamefont {Lakshminarayanan}\ \emph {et~al.}(2017)\citenamefont
  {Lakshminarayanan}, \citenamefont {Pritzel},\ and\ \citenamefont
  {Blundell}}]{lakshminarayanan_simple_2017}%
  \BibitemOpen
  \bibfield  {author} {\bibinfo {author} {\bibfnamefont {B.}~\bibnamefont
  {Lakshminarayanan}}, \bibinfo {author} {\bibfnamefont {A.}~\bibnamefont
  {Pritzel}},\ and\ \bibinfo {author} {\bibfnamefont {C.}~\bibnamefont
  {Blundell}},\ }\bibfield  {title} {\bibinfo {title} {Simple and scalable
  predictive uncertainty estimation using deep ensembles},\ }in\ \href@noop {}
  {\emph {\bibinfo {booktitle} {Advances in Neural Information Processing
  Systems}}},\ Vol.~\bibinfo {volume} {30},\ \bibinfo {editor} {edited by\
  \bibinfo {editor} {\bibfnamefont {I.}~\bibnamefont {Guyon}}, \bibinfo
  {editor} {\bibfnamefont {U.~V.}\ \bibnamefont {Luxburg}}, \bibinfo {editor}
  {\bibfnamefont {S.}~\bibnamefont {Bengio}}, \bibinfo {editor} {\bibfnamefont
  {H.}~\bibnamefont {Wallach}}, \bibinfo {editor} {\bibfnamefont
  {R.}~\bibnamefont {Fergus}}, \bibinfo {editor} {\bibfnamefont
  {S.}~\bibnamefont {Vishwanathan}},\ and\ \bibinfo {editor} {\bibfnamefont
  {R.}~\bibnamefont {Garnett}}}\ (\bibinfo  {publisher} {{Curran Associates,
  Inc.}},\ \bibinfo {year} {2017})\ pp.\ \bibinfo {pages}
  {6402--6413}\BibitemShut {NoStop}%
\bibitem [{\citenamefont {Ovadia}\ \emph {et~al.}(2019)\citenamefont {Ovadia},
  \citenamefont {Fertig}, \citenamefont {Ren}, \citenamefont {Nado},
  \citenamefont {Sculley}, \citenamefont {Nowozin}, \citenamefont {Dillon},
  \citenamefont {Lakshminarayanan},\ and\ \citenamefont
  {Snoek}}]{NEURIPS2019_8558cb40}%
  \BibitemOpen
  \bibfield  {author} {\bibinfo {author} {\bibfnamefont {Y.}~\bibnamefont
  {Ovadia}}, \bibinfo {author} {\bibfnamefont {E.}~\bibnamefont {Fertig}},
  \bibinfo {author} {\bibfnamefont {J.}~\bibnamefont {Ren}}, \bibinfo {author}
  {\bibfnamefont {Z.}~\bibnamefont {Nado}}, \bibinfo {author} {\bibfnamefont
  {D.}~\bibnamefont {Sculley}}, \bibinfo {author} {\bibfnamefont
  {S.}~\bibnamefont {Nowozin}}, \bibinfo {author} {\bibfnamefont
  {J.}~\bibnamefont {Dillon}}, \bibinfo {author} {\bibfnamefont
  {B.}~\bibnamefont {Lakshminarayanan}},\ and\ \bibinfo {author} {\bibfnamefont
  {J.}~\bibnamefont {Snoek}},\ }\bibfield  {title} {\bibinfo {title} {Can you
  trust your models uncertainty? {{Evaluating}} predictive uncertainty under
  dataset shift},\ }in\ \href@noop {} {\emph {\bibinfo {booktitle} {Advances in
  Neural Information Processing Systems}}},\ Vol.~\bibinfo {volume} {32},\
  \bibinfo {editor} {edited by\ \bibinfo {editor} {\bibfnamefont
  {H.}~\bibnamefont {Wallach}}, \bibinfo {editor} {\bibfnamefont
  {H.}~\bibnamefont {Larochelle}}, \bibinfo {editor} {\bibfnamefont
  {A.}~\bibnamefont {Beygelzimer}}, \bibinfo {editor} {\bibfnamefont
  {F.}~\bibnamefont {{dAlch{\'e}-Buc}}}, \bibinfo {editor} {\bibfnamefont
  {E.}~\bibnamefont {Fox}},\ and\ \bibinfo {editor} {\bibfnamefont
  {R.}~\bibnamefont {Garnett}}}\ (\bibinfo  {publisher} {{Curran Associates,
  Inc.}},\ \bibinfo {year} {2019})\BibitemShut {NoStop}%
\bibitem [{\citenamefont {Bl{\"o}chl}(1994)}]{blochl_projector_1994}%
  \BibitemOpen
  \bibfield  {author} {\bibinfo {author} {\bibfnamefont {P.~E.}\ \bibnamefont
  {Bl{\"o}chl}},\ }\bibfield  {title} {\bibinfo {title} {Projector
  augmented-wave method},\ }\href {https://doi.org/10.1103/PhysRevB.50.17953}
  {\bibfield  {journal} {\bibinfo  {journal} {Physical Review B}\ }\textbf
  {\bibinfo {volume} {50}},\ \bibinfo {pages} {17953} (\bibinfo {year}
  {1994})}\BibitemShut {NoStop}%
\bibitem [{\citenamefont {Perdew}\ \emph {et~al.}(1996)\citenamefont {Perdew},
  \citenamefont {Burke},\ and\ \citenamefont
  {Ernzerhof}}]{perdew_generalized_1996}%
  \BibitemOpen
  \bibfield  {author} {\bibinfo {author} {\bibfnamefont {J.~P.}\ \bibnamefont
  {Perdew}}, \bibinfo {author} {\bibfnamefont {K.}~\bibnamefont {Burke}},\ and\
  \bibinfo {author} {\bibfnamefont {M.}~\bibnamefont {Ernzerhof}},\ }\bibfield
  {title} {\bibinfo {title} {Generalized {{Gradient Approximation Made
  Simple}}},\ }\href {https://doi.org/10.1103/PhysRevLett.77.3865} {\bibfield
  {journal} {\bibinfo  {journal} {Physical Review Letters}\ }\textbf {\bibinfo
  {volume} {77}},\ \bibinfo {pages} {3865} (\bibinfo {year}
  {1996})}\BibitemShut {NoStop}%
\bibitem [{\citenamefont {Kresse}\ and\ \citenamefont
  {Furthm{\"u}ller}(1996{\natexlab{a}})}]{kresse_efficiency_1996}%
  \BibitemOpen
  \bibfield  {author} {\bibinfo {author} {\bibfnamefont {G.}~\bibnamefont
  {Kresse}}\ and\ \bibinfo {author} {\bibfnamefont {J.}~\bibnamefont
  {Furthm{\"u}ller}},\ }\bibfield  {title} {\bibinfo {title} {Efficiency of
  ab-initio total energy calculations for metals and semiconductors using a
  plane-wave basis set},\ }\href {https://doi.org/10.1016/0927-0256(96)00008-0}
  {\bibfield  {journal} {\bibinfo  {journal} {Computational Materials Science}\
  }\textbf {\bibinfo {volume} {6}},\ \bibinfo {pages} {15} (\bibinfo {year}
  {1996}{\natexlab{a}})}\BibitemShut {NoStop}%
\bibitem [{\citenamefont {Kresse}\ and\ \citenamefont
  {Furthm{\"u}ller}(1996{\natexlab{b}})}]{kresse_efficient_1996}%
  \BibitemOpen
  \bibfield  {author} {\bibinfo {author} {\bibfnamefont {G.}~\bibnamefont
  {Kresse}}\ and\ \bibinfo {author} {\bibfnamefont {J.}~\bibnamefont
  {Furthm{\"u}ller}},\ }\bibfield  {title} {\bibinfo {title} {Efficient
  iterative schemes for ab initio total-energy calculations using a plane-wave
  basis set},\ }\href {https://doi.org/10.1103/PhysRevB.54.11169} {\bibfield
  {journal} {\bibinfo  {journal} {Physical Review B}\ }\textbf {\bibinfo
  {volume} {54}},\ \bibinfo {pages} {11169} (\bibinfo {year}
  {1996}{\natexlab{b}})}\BibitemShut {NoStop}%
\bibitem [{Note1()}]{Note1}%
  \BibitemOpen
  \bibinfo {note} {The code and the pre-trained model are from \protect \url
  {https://github.com/txie-93/cgcnn}}\BibitemShut {NoStop}%
\bibitem [{\citenamefont {Yosinski}\ \emph {et~al.}(2014)\citenamefont
  {Yosinski}, \citenamefont {Clune}, \citenamefont {Bengio},\ and\
  \citenamefont {Lipson}}]{yosinski_how_2014}%
  \BibitemOpen
  \bibfield  {author} {\bibinfo {author} {\bibfnamefont {J.}~\bibnamefont
  {Yosinski}}, \bibinfo {author} {\bibfnamefont {J.}~\bibnamefont {Clune}},
  \bibinfo {author} {\bibfnamefont {Y.}~\bibnamefont {Bengio}},\ and\ \bibinfo
  {author} {\bibfnamefont {H.}~\bibnamefont {Lipson}},\ }\bibfield  {title}
  {\bibinfo {title} {How transferable are features in deep neural networks?},\
  }in\ \href@noop {} {\emph {\bibinfo {booktitle} {Advances in Neural
  Information Processing Systems}}},\ Vol.~\bibinfo {volume} {27},\ \bibinfo
  {editor} {edited by\ \bibinfo {editor} {\bibfnamefont {Z.}~\bibnamefont
  {Ghahramani}}, \bibinfo {editor} {\bibfnamefont {M.}~\bibnamefont {Welling}},
  \bibinfo {editor} {\bibfnamefont {C.}~\bibnamefont {Cortes}}, \bibinfo
  {editor} {\bibfnamefont {N.}~\bibnamefont {Lawrence}},\ and\ \bibinfo
  {editor} {\bibfnamefont {K.~Q.}\ \bibnamefont {Weinberger}}}\ (\bibinfo
  {publisher} {{Curran Associates, Inc.}},\ \bibinfo {year} {2014})\ pp.\
  \bibinfo {pages} {3320--3328}\BibitemShut {NoStop}%
\bibitem [{\citenamefont {Kingma}\ and\ \citenamefont
  {Ba}(2017)}]{kingma_adam_2017}%
  \BibitemOpen
  \bibfield  {author} {\bibinfo {author} {\bibfnamefont {D.~P.}\ \bibnamefont
  {Kingma}}\ and\ \bibinfo {author} {\bibfnamefont {J.}~\bibnamefont {Ba}},\
  }\bibfield  {title} {\bibinfo {title} {Adam: {{A Method}} for {{Stochastic
  Optimization}}},\ }\href@noop {} {\bibfield  {journal} {\bibinfo  {journal}
  {arXiv:1412.6980 [cs]}\ } (\bibinfo {year} {2017})},\ \Eprint
  {https://arxiv.org/abs/1412.6980} {arXiv:1412.6980 [cs]} \BibitemShut
  {NoStop}%
\bibitem [{Note2()}]{Note2}%
  \BibitemOpen
  \bibinfo {note} {See Supplemental Material at [URL will be inserted by
  publisher] for full details on the results}\BibitemShut {NoStop}%
\bibitem [{\citenamefont {Ong}\ \emph {et~al.}(2013)\citenamefont {Ong},
  \citenamefont {Richards}, \citenamefont {Jain}, \citenamefont {Hautier},
  \citenamefont {Kocher}, \citenamefont {Cholia}, \citenamefont {Gunter},
  \citenamefont {Chevrier}, \citenamefont {Persson},\ and\ \citenamefont
  {Ceder}}]{ong_python_2013}%
  \BibitemOpen
  \bibfield  {author} {\bibinfo {author} {\bibfnamefont {S.~P.}\ \bibnamefont
  {Ong}}, \bibinfo {author} {\bibfnamefont {W.~D.}\ \bibnamefont {Richards}},
  \bibinfo {author} {\bibfnamefont {A.}~\bibnamefont {Jain}}, \bibinfo {author}
  {\bibfnamefont {G.}~\bibnamefont {Hautier}}, \bibinfo {author} {\bibfnamefont
  {M.}~\bibnamefont {Kocher}}, \bibinfo {author} {\bibfnamefont
  {S.}~\bibnamefont {Cholia}}, \bibinfo {author} {\bibfnamefont
  {D.}~\bibnamefont {Gunter}}, \bibinfo {author} {\bibfnamefont {V.~L.}\
  \bibnamefont {Chevrier}}, \bibinfo {author} {\bibfnamefont {K.~A.}\
  \bibnamefont {Persson}},\ and\ \bibinfo {author} {\bibfnamefont
  {G.}~\bibnamefont {Ceder}},\ }\bibfield  {title} {\bibinfo {title} {Python
  {{Materials Genomics}} (pymatgen): {{A}} robust, open-source python library
  for materials analysis},\ }\href
  {https://doi.org/10.1016/j.commatsci.2012.10.028} {\bibfield  {journal}
  {\bibinfo  {journal} {Computational Materials Science}\ }\textbf {\bibinfo
  {volume} {68}},\ \bibinfo {pages} {314} (\bibinfo {year} {2013})}\BibitemShut
  {NoStop}%
\end{thebibliography}
\end{document}